\documentclass{IEEEtran}
\usepackage{cite}
\usepackage{amssymb,amsfonts}
\usepackage{amsmath}
\usepackage{algorithm}
\usepackage{algpseudocode} % Use algpseudocode instead of algorithmic
\usepackage{textcomp}
\usepackage{graphicx}%
\usepackage{multirow}%
\usepackage{mathrsfs}%
\usepackage{manyfoot}%
\usepackage{booktabs}%
\usepackage{multicol}
\usepackage{listings}%
\usepackage{subcaption}
\usepackage{acronym} % used for managing acronyms

\acrodef{RCTM}{Robust Chaotic Tent Map}
\acrodef{PRNG}{Pseudo-random Number Generator}
\acrodef{CSPRNG}{Cryptographically Secure Pseudo-random Number Generator}

\def\BibTeX{{\rm B\kern-.05em{\sc i\kern-.025em b}\kern-.08em
    T\kern-.1667em\lower.7ex\hbox{E}\kern-.125emX}}
\begin{document}
\title{Cryptographically Secure Pseudo-Random Number Generation (CS-PRNG) Design using Robust Chaotic Tent Map (RCTM)}
\author{Muhammad Irfan and Muhammad {Asif Khan}
%\thanks{Open Access funding provided by the Qatar National Library. }

\thanks{M. Irfan is with Division of Information and Computing Technology, College of Science and Engineering, Hamad Bin Khalifa University, Qatar Foundation - Doha, Qatar (Corresponding Author e-mail: muir45306@hbku.edu.qa)}
\thanks{M. A. Khan is with 
Department of Computer Engineering, University of Engineering and Technology Taxila 47050, Pakistan (e-mail: masif.khan@uettaxila.edu.pk).}}

\maketitle

\begin{abstract}
Chaos, a nonlinear dynamical system, favors cryptography due to their inherent sensitive dependence on the initial condition, mixing, and ergodicity property. In recent years, the nonlinear behavior of chaotic maps has been utilized as a random source to generate pseudo-random number generation for cryptographic services. For chaotic maps having Robust chaos, dense, chaotic orbits exist for the range of parameter space—the occurrence of chaotic attractors in some neighborhoods of parameter space and the absence of periodic windows. Thus, the robust chaotic map shows assertive chaotic behavior for larger parameters space with a positive Lyapunov exponent. This paper presents a novel method to generate cryptographically secure pseudo-random numbers (CSPRNG) using a robust chaotic tent map (RCTM). We proposed a new set of equations featuring modulo and scaling operators that achieve vast parameter space by keeping chaotic orbit globally stable and robust. The dynamic behavior of the RCTM is studied first by plotting the bifurcation diagram that shows chaotic behavior for different parameters, which the positive Lyapunov exponent verifies. We iterated the RCTM to generate pseudo-random bits using a simple thresholding method. Various statistical tests are performed that ascertain the randomness of generated secure pseudo-random bits. It includes NIST 800-22 test suite, ENT statistical test suite, TestU01 test suite, key space analysis, key sensitivity analysis, correlation analysis, histogram analysis, and differential analysis. The proposed scheme has achieved larger key space as compared with existing methods. The results show that the proposed PRBG algorithm can generate CSPRNG.
\end{abstract}


\section{Introduction}
\label{sec:introduction}
As the world increasingly shifts to digital platforms, securing data effectively has become more challenging. To address this, cryptographic mechanisms such as hash functions, encryption, digital signatures, and authentication protocols for data and devices, as well as message authentication codes, have been developed. These mechanisms often incorporate \ac{PRNG} in their design for critical tasks like key generation, one-time passwords (OTP), and initialization vectors (IV). \ac{PRNG}s are utilized extensively across a variety of applications, including cryptographic systems~\cite{bai2019digital,soujeri2018design}, spread spectrum communications~\cite{hull1962random}, statistics~\cite{eisencraft2019new}, signal processing~\cite{alani2019applications,taran2018bridging}, machine learning~\cite{meoni1997casino,david1998Central}, gaming~\cite{waqas2018stochastic,ghauch2019integrated}, and stochastic processes. PRNGs are classified into two types: deterministic and non-deterministic. Deterministic PRNGs, which generate numbers based on a seed, can be reproduced. Conversely, non-deterministic PRNGs, also known as true random number generators (TRNGs), rely on natural and physical phenomena and are non-reproducible. Despite their advantage of being statistically unbiased and unpredictable~\cite{schindler2003evaluation}, TRNGs suffer from the disadvantage of slow true random number generation.

On the other hand, \ac{PRNG}s are mathematical functions triggered by an initial condition; therefore, they are deterministic functions that can be fully determined using the same initial condition originally employed to produce the \ac{PRNG}. The properties of reproducibility, repeatability, and fast generation make PRNGs more vulnerable than TRNGs. A specific type of \ac{PRNG}, known as a \ac{CSPRNG}, is required in cryptographic applications. \ac{CSPRNG}s are highly nonlinear and unpredictable, features that are essential for the design of cryptographic services.

Over the past two decades, the use of chaotic maps in cryptography has significantly increased. Cryptography and chaos theory share several common features, including high unpredictability, ergodic behavior, extreme sensitivity to initial conditions, deterministic dynamics, and complex behavior. When implemented using a finite state machine, chaotic maps are referred to as chaotic digital maps. The control parameters and initial conditions, which serve as the seeds for the chaotic system, define the characteristics of chaotic trajectories. It is crucial to select parameters that promote near-optimal randomness~\cite{demir2018analysis}. These digital chaotic maps are extensively employed in chaos-based cryptographic algorithms to enhance security through confusion~\cite{shannon1949communication,teh2017chaos,teh2019chaos,teh2015parallel}. The trajectories generated by iterating chaotic maps are inherently nonlinear. This nonlinear and unpredictable behavior, effectively measured by the Lyapunov exponent, can be preserved during binary mapping by applying methods such as thresholding carefully.

It has been observed recently, chaotic maps such as logistic~\cite{wang2019pseudo}, henon map, skew-tent map, and sine maps have been used in various cryptographic systems. Their implementation has significant limitations, such as lower parameters range, susceptibility to attacks~\cite{ho2010parameter,arroyo2010inadequacy}, and shorter periods. The nature of chaotic behavior is inherited for chaotic maps. However, when subjected to digital form, they face limitations in the small size of bits which causes chaotic behavior to be degraded, and the trajectory may approximate periodically. These problems are indispensable to be discussed before prevailing chaotic maps are utilized in cryptographic systems. Different techniques are proposed to overcome these weaknesses by applying genetic algorithms, cellular automata, a combination of other maps, etc.

The structure of one-dimensional (1D) chaotic maps makes their implementation simple; however, it jeopardizes parameter estimation attacks. Several researchers have concentrated on the high dimensional (HD) chaotic maps and claim better security than their rival 1D chaotic maps~\cite{hua20152d}. Some authors have used HD chaotic maps in a few algorithms without the security improvements~\cite{seyedzadeh2012fast,chen2004symmetric,wang2013secure}. However, the security of HD chaotic maps is compromised by cryptanalysis attacks~\cite{wang2013secure,pisarchik2008image}.

Various researchers analyzed the design and implementation of \ac{PRNG}, both in hardware and software. Moysis et al.~\cite{moysis2021chaotic} proposed a novel chaotic map, they introduced a 1D map with a simple structure and three parameters. The proposed scheme generated four bits in each iteration to produce bitstream at a faster rate. The statistical analysis is carried out with the National Institute of Standards and Technology (NIST) 800-22 statistical test suite (STS)~\cite{bassham2010sp} and ENT~\cite{WalkerPseudorandom} stamped its random behavior. Krishnamoorthy et al.~\cite{krishnamoorthi2021design} familiarized the logistic map with turbulence padded chaotic map to increase the period and chaotic behavior. The devised PRNG architecture meets the minimum criteria of randomness. Irfan et al.~\cite{irfan2020pseudorandom}, author of this paper, introduced control of Chaos (CoC), where the modified logistic map is used with a larger control parameter space. The CoC technique employed the robust chaotic logistic map to generate \ac{CSPRNG}. The proposed method was also implemented on the various chaotic maps, such as Circle map, Iterative map, Tent map, and Singer maps. The NIST has been successfully implemented on these maps entails CSPRNG. Wang et al.~\cite{wang2016pseudorandom} proposed a piecewise logistic map to enhance the range of control parameters. The generated binary sequence was developed and tested for \ac{CSPRNG} using the NIST test suite. The logistic map is modified by applying modular operation and adding a multiplier. The enhanced map is suitable for PRNG in the embedded system. Zia et. al.~\cite{zia2022novel} leveraged coupled map lattice to generate PRNG for Internet of Things (IoT) devices, they tested their design with Raspberry Pi's variants (zeroW, 3B+). Authors in~\cite{zia2023resource} used SPONGENT hash function and sawtooth couple map lattice to generate PRNG for IoT devices, they tested their design on Rasberry Pi zeroW, 3B+ and 4.

Reconfigurable hardware-based \ac{PRNG} is proposed in~\cite{rezk2019reconfigurable}, which is based on an HD chaotic map. The Lorenz~\cite{lorenz1963deterministic} and Lü~\cite{lu2002new} chaotic maps are used to produce three-dimensional chaotic attractors. The proposed method is reconfigured and synthesized on a field-programmable array FPGA. The NIST STS is performed to observe the behavior of generated random sequence. Rania and Ehab~\cite{elmanfaloty2019random} examined the study of fixed-point notation of random numbers and their outcomes on randomness and periodicity. The authors proposed a hardware-based PRNG, utilizing multiple maps of chaotic skew tent map~\cite{hasler1997introduction}, cross-coupled chaotic skew ten, map, and couple skew tent map. A hardware-based implementation of Sundarapandian—Pehlivan chaotic system~\cite{sundarapandian2012analysis} is carried out on Virtex-6 FPGA with a data rate of 58.76 Megabits/s by Koyuncu and Özcerit~\cite{koyuncu2017design}. Avaroğlu et al.~\cite{avaroglu2014new}, proposed a hybrid approach to generate \ac{PRNG}. The chaotic system provides an additional input to the sky, stem which is implemented on FPGA. The successful NIST STS proof of unpredictable \ac{PRNG} is presented.  

Robust chaotic maps with positive Lyapunov exponent and large parameter space are desirable for cryptographic applications due to the amenity of choosing control parameter space and initial conditions~\cite{gayathri2016survey}. In this paper, a simple method of CS-PRNG using RCTM is proposed. The initial trajectory is generated using RCTM with arbitrary chosen initial conditions and control parameters. Simple thresholding is applied that maps the trajectory onto a binary sequence. The proposed RCTM is robust for the entire control parameter region [2,100]. The assertive behavior is evident from the positive Lyapunov exponent.

Further, a chaotic attractor exists in the neighborhood of parameter space, and the absence of a periodic window is observed in the bifurcation diagram drawn. Various standard and attested tests are performed, such as NIST 800-22 statistical test suite~\cite{bassham2010sp} and ENT~\cite{WalkerPseudorandom}, histogram analysis, key sensitivity, and key space analysis. The proposed simple threshold-based configuration can generate CS-PRNG. The proposed study can be used effectively in various cryptographic applications. 

This paper is organized into the following sections: Section~\ref{sec:tent_map} and \ref{sec:modified_tent_map} discusses the properties of Chaotic Tent Map and Robust Chaotic Tent Map such as ergodicity, bifurcation, and Lyapunov exponent. Section~\ref{sec:methodology} presents the methodology for generating PRBG. Section~\ref{sec:analysis} presents the performance analysis of the proposed PRBG, including NIST 800-22 test suit, ENT, key-space analysis, correlation behavior of two closely related parameters, information entropy, histogram analysis, and sensitivity to initial conditions. In last section~\ref{sec:conclusion}, the conclusion of the proposed study is given.
\begin{figure}
    \centering
    \includegraphics[trim=0 0 0 22, clip,width=\columnwidth]{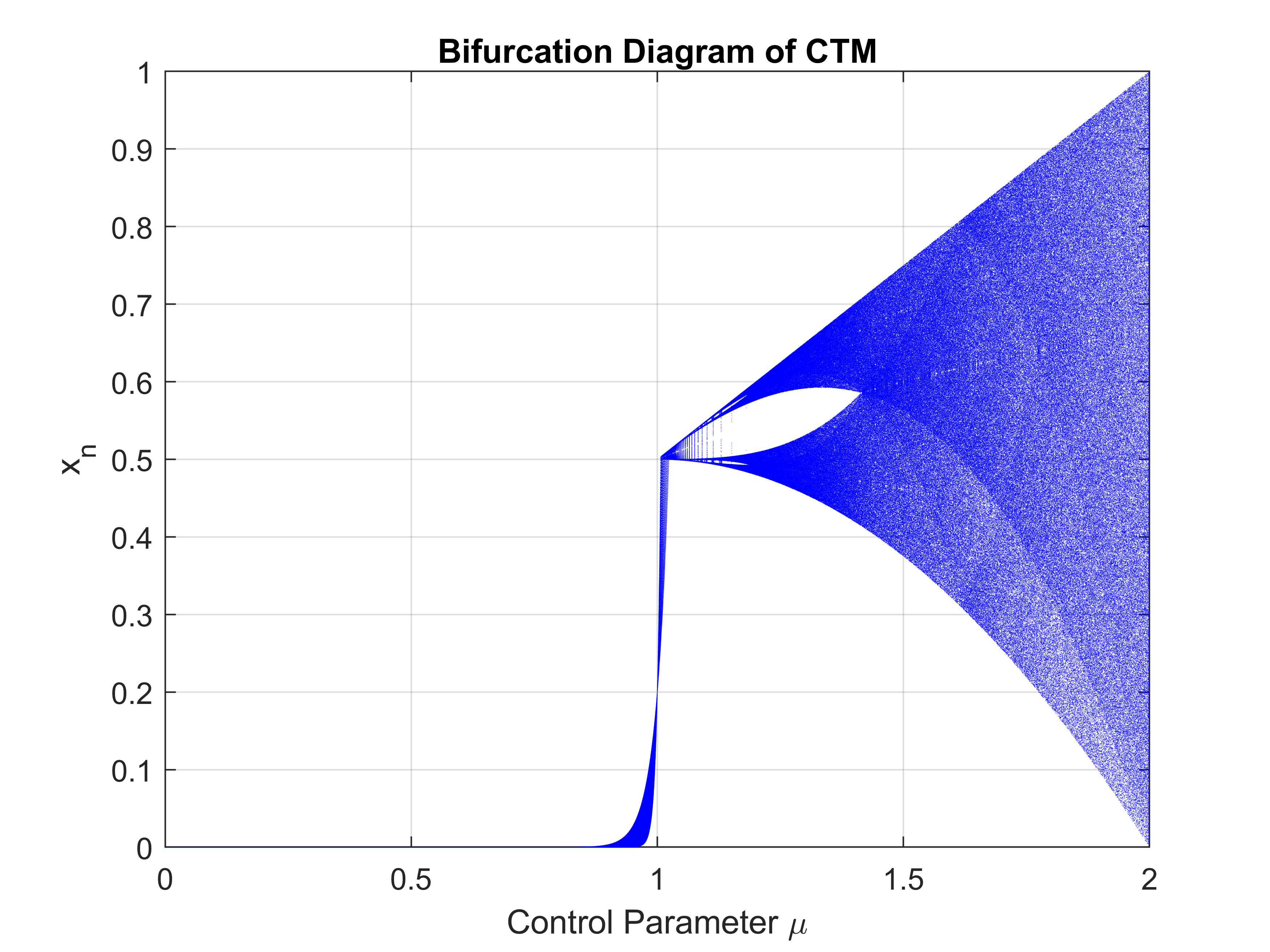}
    \caption{Bifurcation diagram of Classical Tent Map}
    \label{fig:tent_bifurcation}
\end{figure}

\begin{figure}
    \centering
    \includegraphics[trim=0 0 0 22, clip,width=\columnwidth]{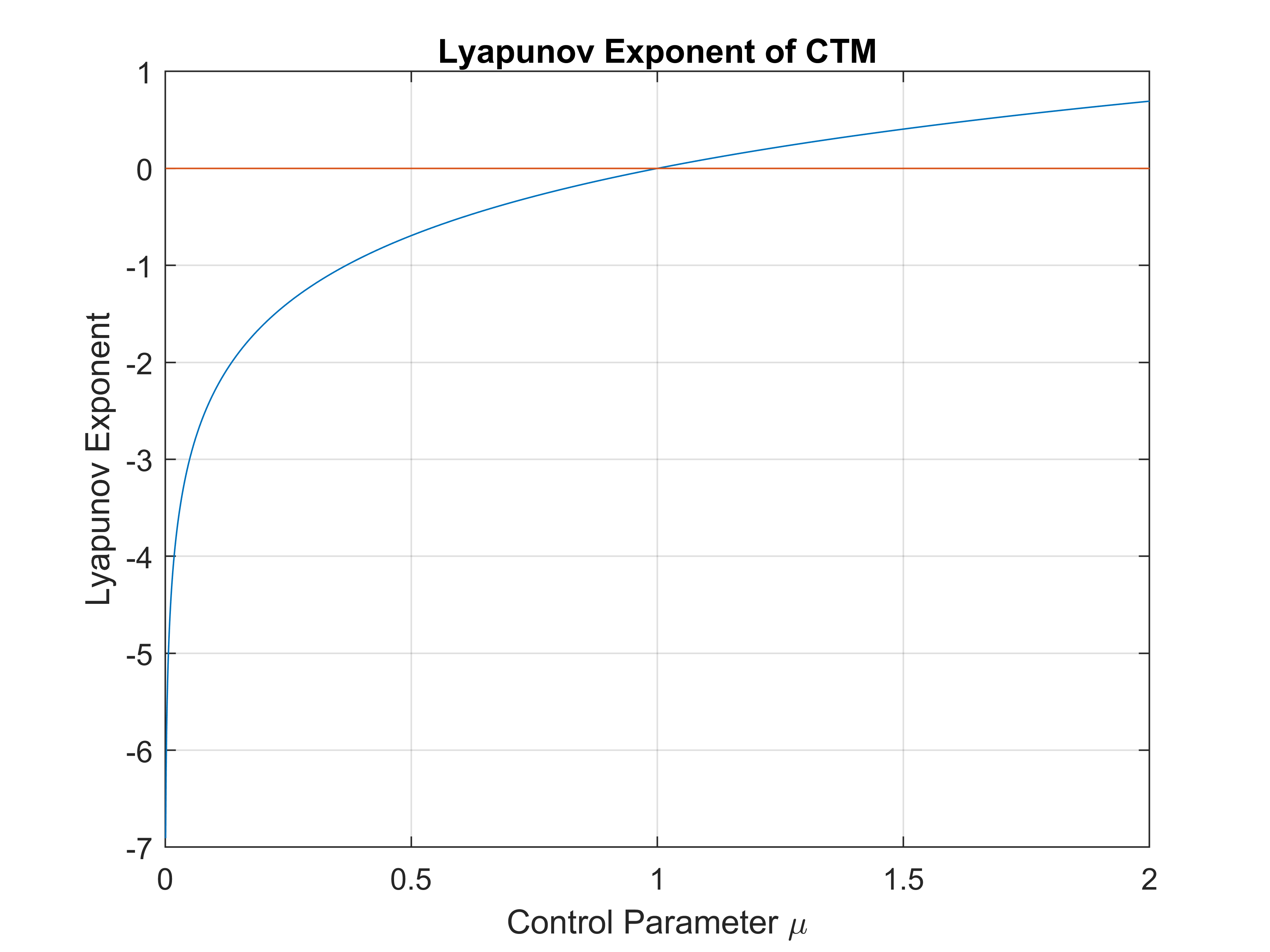}
    \caption{Lyapunov Exponent of Classical Tent Map}
    \label{fig:tent_layp}
\end{figure}

\section{Chaotic Tent Map}
\label{sec:tent_map}

The chaotic tent map (CTM) is a 1D map based on a single mathematical equation. The CTM, a simple structure, is useful for cryptographic applications to generate \ac{PRNG} but limited key parameter space. Eq.~\eqref{eq:tent_map} defines the 1D CTM,

\begin{equation}
x_{n+1} = 
    \begin{cases}
        \mu x_n, & \text{if } x_n < \frac{1}{2}\\
        \mu (1-x_n), & \text{if } x_n \geq \frac{1}{2}
    \end{cases}
\label{eq:tent_map}
\end{equation}

Where $\mu$ is the controlling parameter lies in the range of $[0,2]$ and, $x_n$ is initial condition $\in (0,1)$. 
The CTM is an uneven piecewise chaotic map, which exhibits chaotic behavior at $\mu = 2$. To examine the chaotic range bifurcation diagram is presented in Fig.~\ref{fig:tent_bifurcation}, which shows that the chaotic system covers all space states at $\mu = 2$. The Lyapunov exponent (LE) of CTM is given in Fig.~\ref{fig:tent_layp}. The LE is positive for $\mu \in (1,2)$, but this region is not favorable for cryptographic application as the map covers entire state values only at $\mu = 2$.

\section{Robust Chaotic Tent Map (RCTM)}
\label{sec:modified_tent_map}

The chaotic tent map has been modified to enhance its robustness across a parameter space of $\mu \in [2, 100]$~\cite{khan2013modified}. This revised version, known as the \ac{RCTM}, incorporates scaling and shifting operations. The modified mathematical equations for the \ac{RCTM} are given below:

\begin{align}
    n_1 = \frac{1}{2} - \frac{\left( \mu \frac{1}{2}\right) mod 1}{\mu}
    \label{eq:n1_find}
\end{align}

\begin{align}
    n_2 = \frac{1}{2} + \frac{\left( \mu \frac{1}{2}\right) mod 1}{\mu}
    \label{eq:n2_find}
\end{align}

\begin{equation}
x_{n+1} = 
    \begin{cases}
        \frac{\mu x_n mod 1}{ \left( \frac{\mu}{2} \right) mod 1 }, & \text{if } x_n < \frac{1}{2}\\
         \frac{\left( \mu \left( 1- x_n \right) \right) mod 1}{ \left( \frac{\mu}{2} \right) mod 1 },  & \text{if } x_n \geq \frac{1}{2}
    \end{cases}
\label{eq:modify_tent_map1}
\end{equation}

\begin{equation}
x_{n+1} = 
    \begin{cases}
        \mu x_n mod 1, & \text{if } x_n < \frac{1}{2}\\
        \mu (1-x_n) mod 1, & \text{if } x_n \geq \frac{1}{2}
    \end{cases}
\label{eq:modify_tent_map2}
\end{equation}

\begin{figure}
  \centering
  \begin{subfigure}{\columnwidth}
    \includegraphics[width=\linewidth]{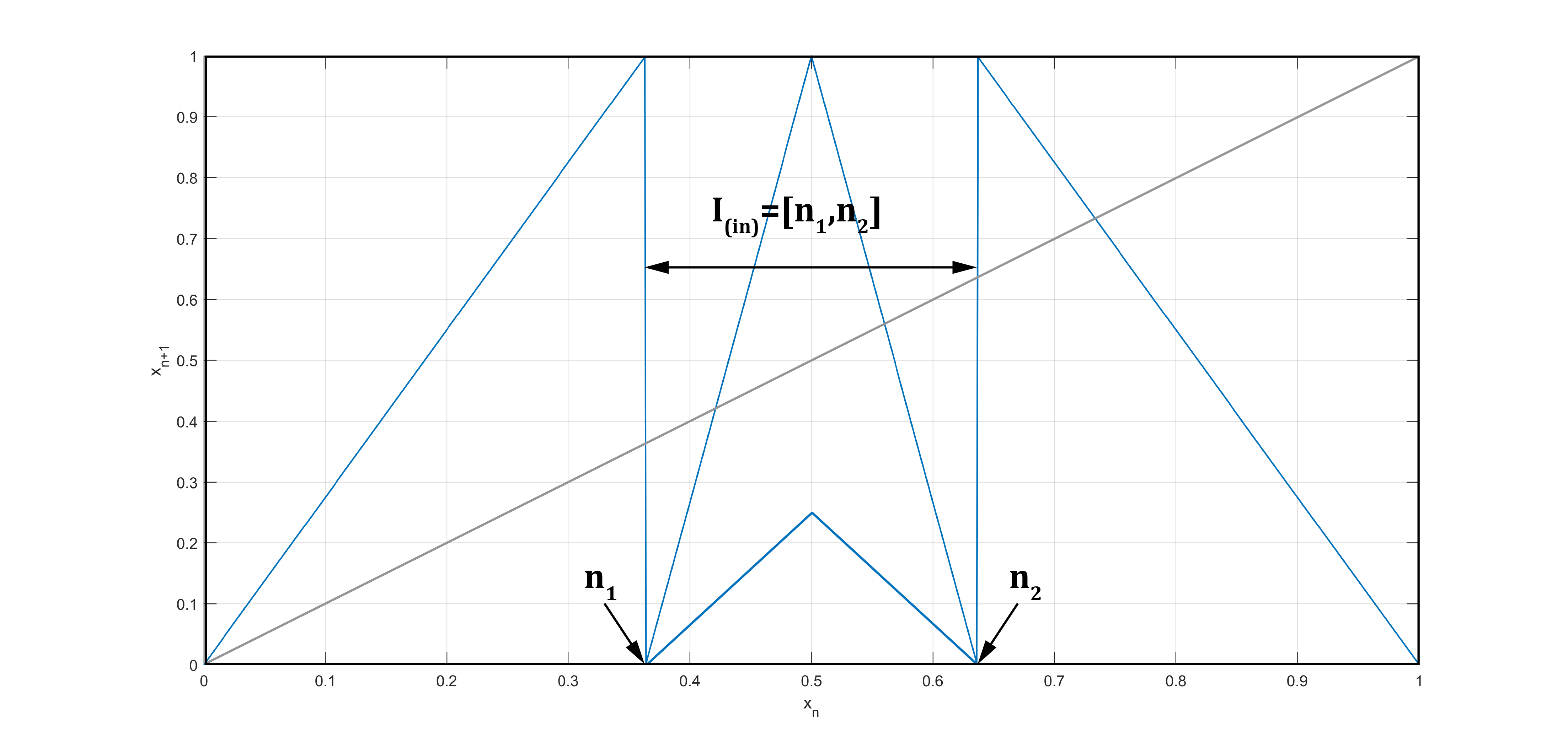}
    \caption{}
    \label{fig:figure_a}
  \end{subfigure}
  \hspace{0.05\textwidth}
  \begin{subfigure}{\columnwidth}
    \includegraphics[width=\linewidth]{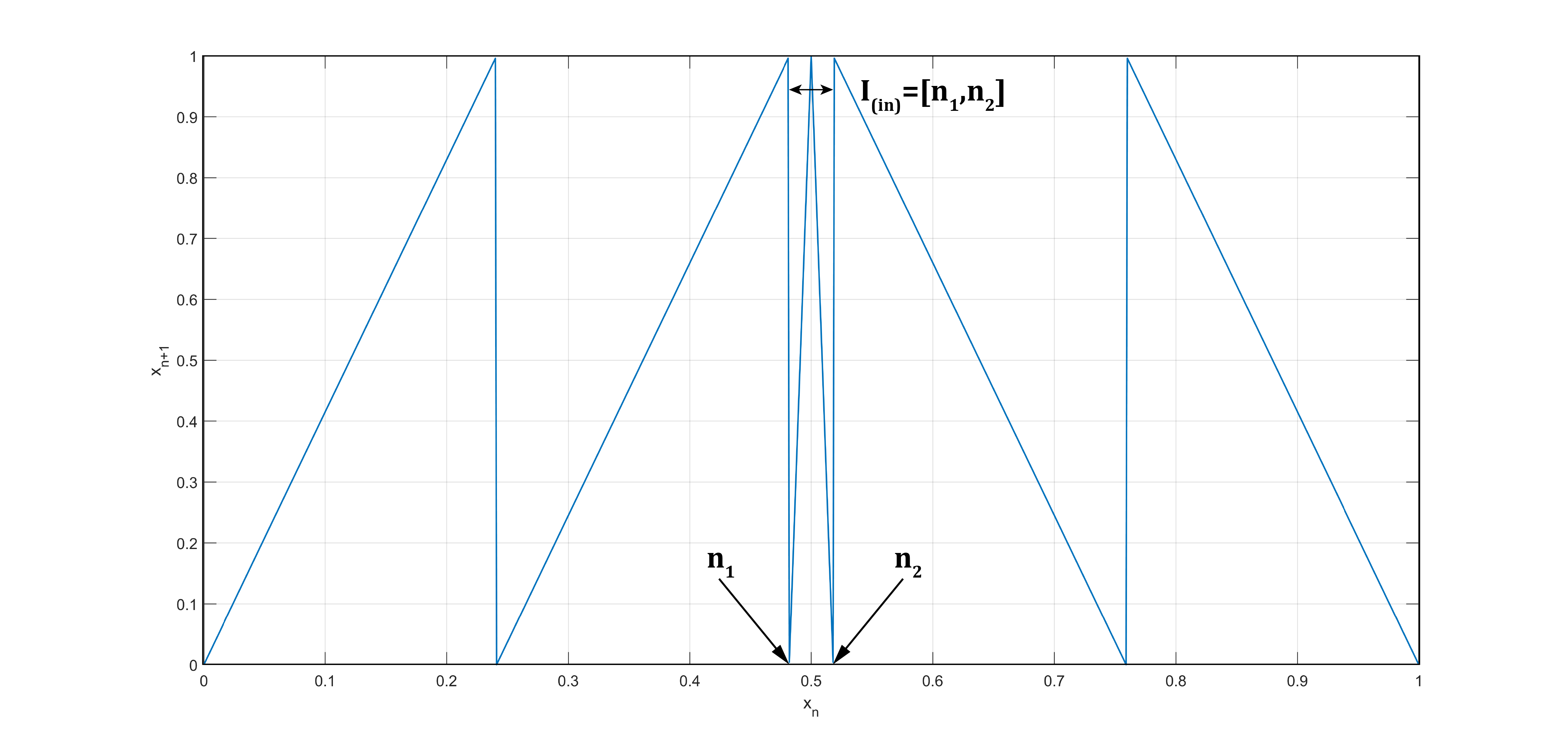}
    \caption{}
    \label{fig:figure_b}
  \end{subfigure}
  \caption{RCTM with Scaling and modulo operation a) $\mu=2.75$ b) $\mu= 4.15$ }
  \label{fig:scaling_modolu_operation}
\end{figure}

\begin{figure}
    \centering
    \includegraphics[width=\columnwidth]{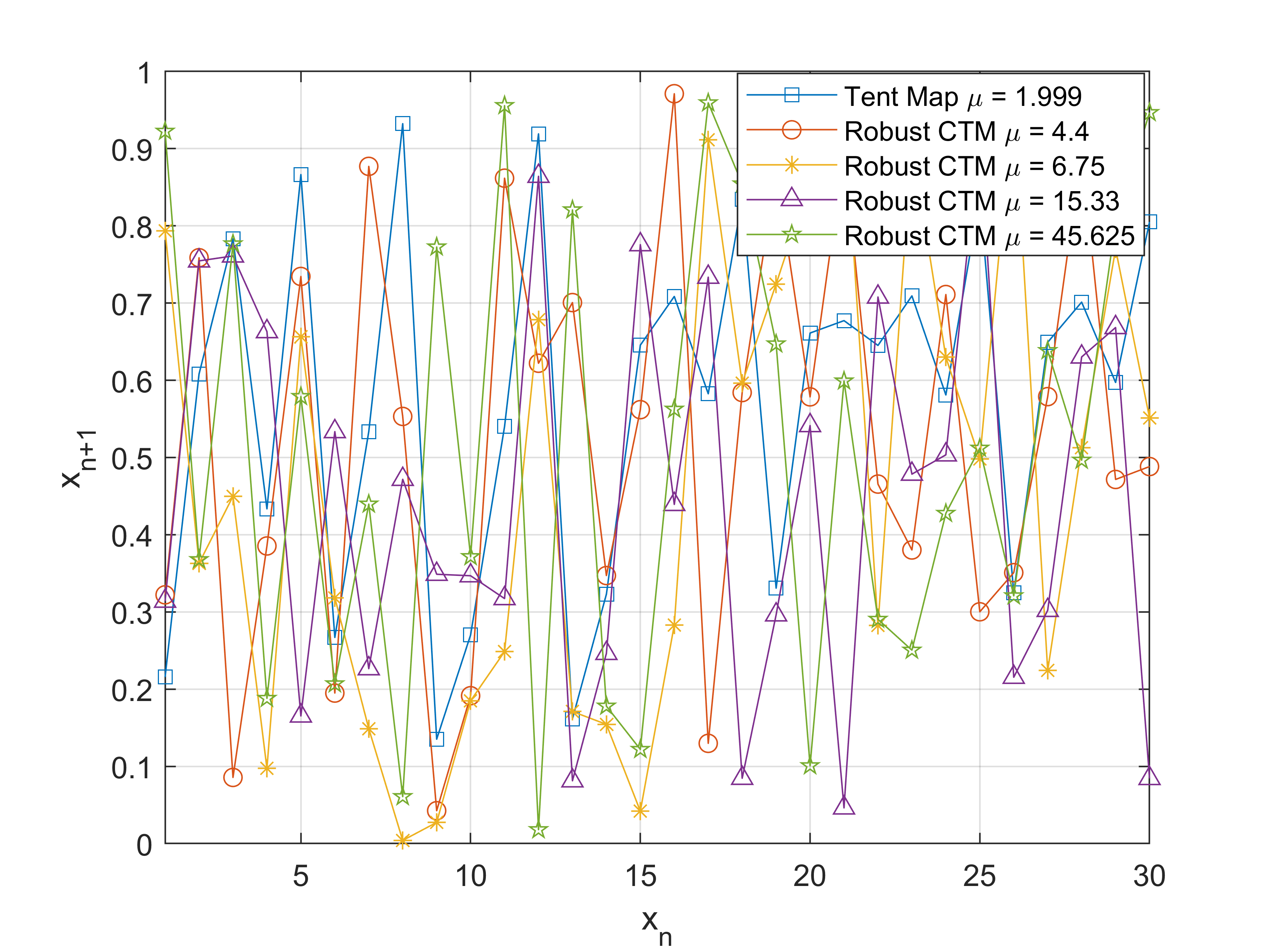}
    \caption{Correlation between the generated sequences of Tent Map and CTM with variable control parameter}
    \label{fig:in_out_sequence}
\end{figure}

The above series of Eqs.~\eqref{eq:n1_find}~\eqref{eq:n2_find}~\eqref{eq:modify_tent_map1}~\eqref{eq:modify_tent_map2} defines the \ac{RCTM}. The domain of \ac{RCTM} is divided into two subspace, $I_{in} \in [n_1,n_2]$ and $I_{ext} \in (0,1) \setminus I_{in}$. Eq.~\eqref{eq:modify_tent_map1} is selected when $x_n \in I_{in}$ and Eq.~\eqref{eq:modify_tent_map1} has opted when $x_n \in I_{ext}$. The Fig.~\ref{fig:scaling_modolu_operation} presents the modulo and scaling operations used in~\eqref{eq:modify_tent_map1}~\eqref{eq:modify_tent_map2}. The CTM can generate chaotic trajectories with arbitrary initial conditions and $\mu=2$. Whereas,  with chaotic attractors and no periodic windows, the RCTM can generate chaotic trajectories with an arbitrary initial condition and control parameter $\mu \in [2,100]$. The correlation among trajectories generated using CTM and RCTM can be observed in Fig.~\ref{fig:in_out_sequence}. The trajectories generated by CTM and RCTM are highly decorrelated, and even small perturbations in the initial condition can lead to completely different trajectories. 

\begin{figure*}
  \centering
  \begin{subfigure}{0.24\textwidth}
    \includegraphics[trim=0 0 0 22, clip, width=\linewidth]{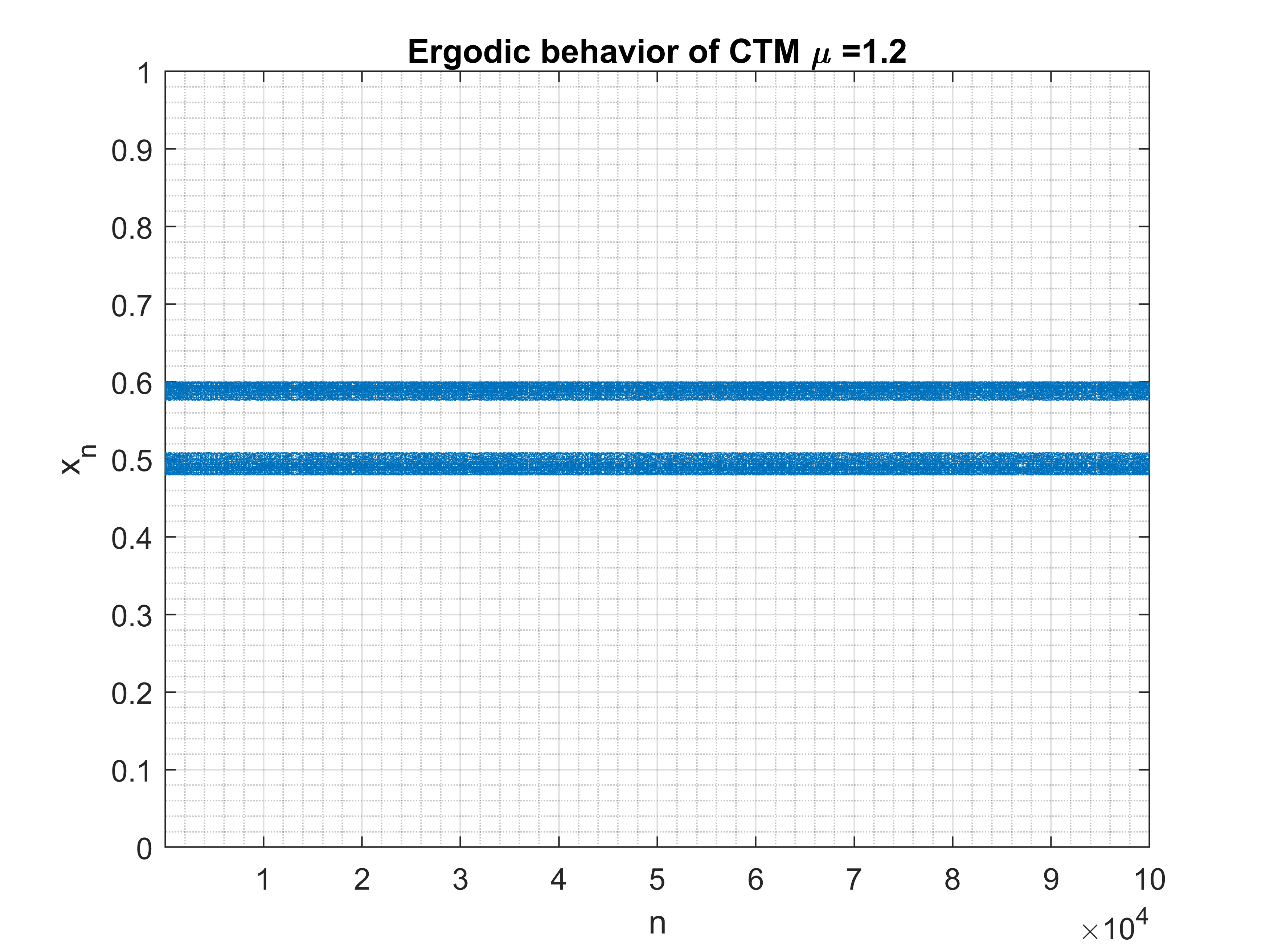}
    \caption{$\mu = 1.2$}
    \label{fig:image1}
  \end{subfigure}
  \hfill
  \begin{subfigure}{0.24\textwidth}
    \includegraphics[trim= 0 0 0 22, clip, width=\linewidth]{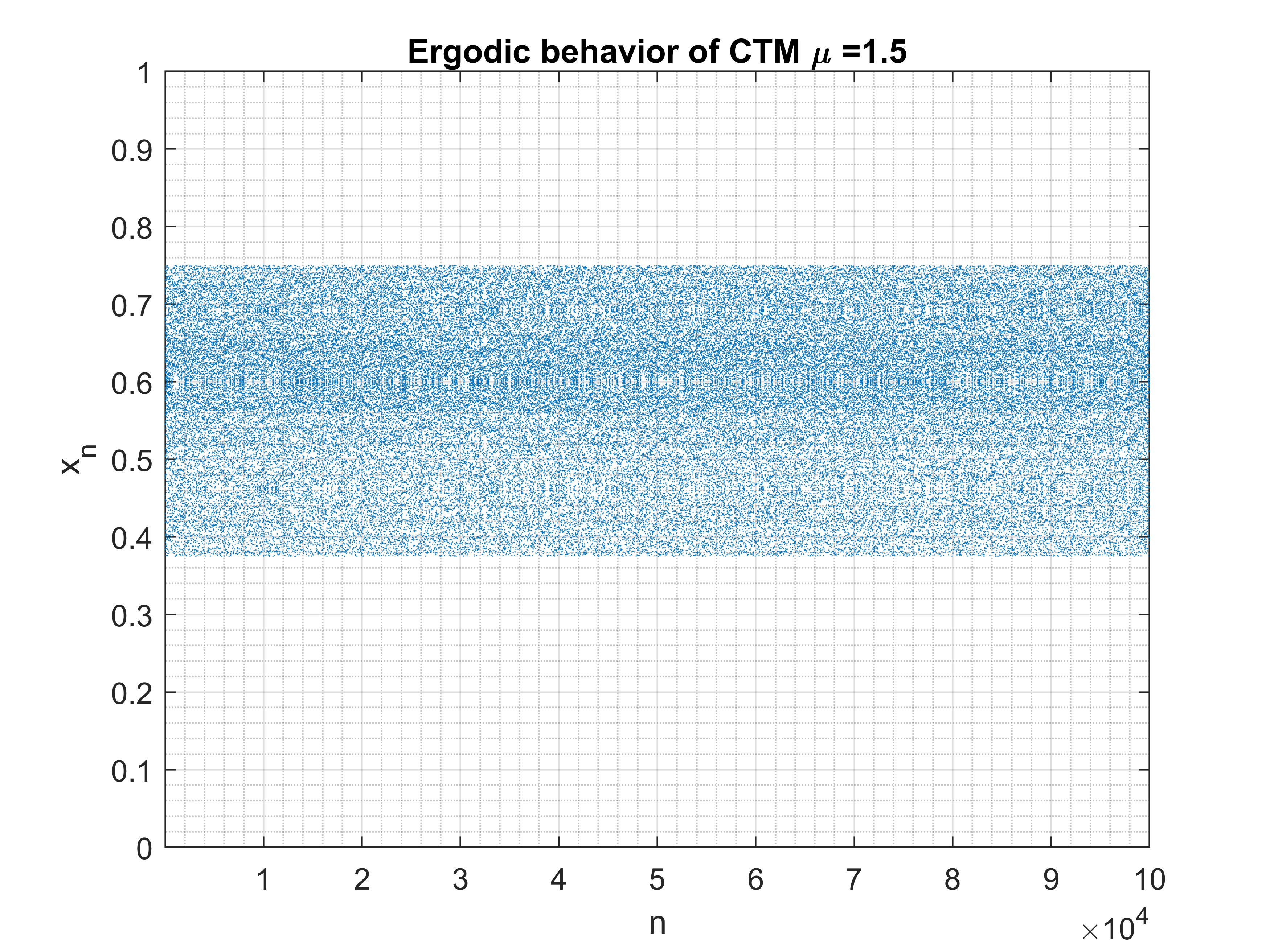}
    \caption{$\mu = 1.5$}
    \label{fig:image2}
  \end{subfigure}
  \hfill
  \begin{subfigure}{0.24\textwidth}
    \includegraphics[trim=0 0 0 22, clip, width=\linewidth]{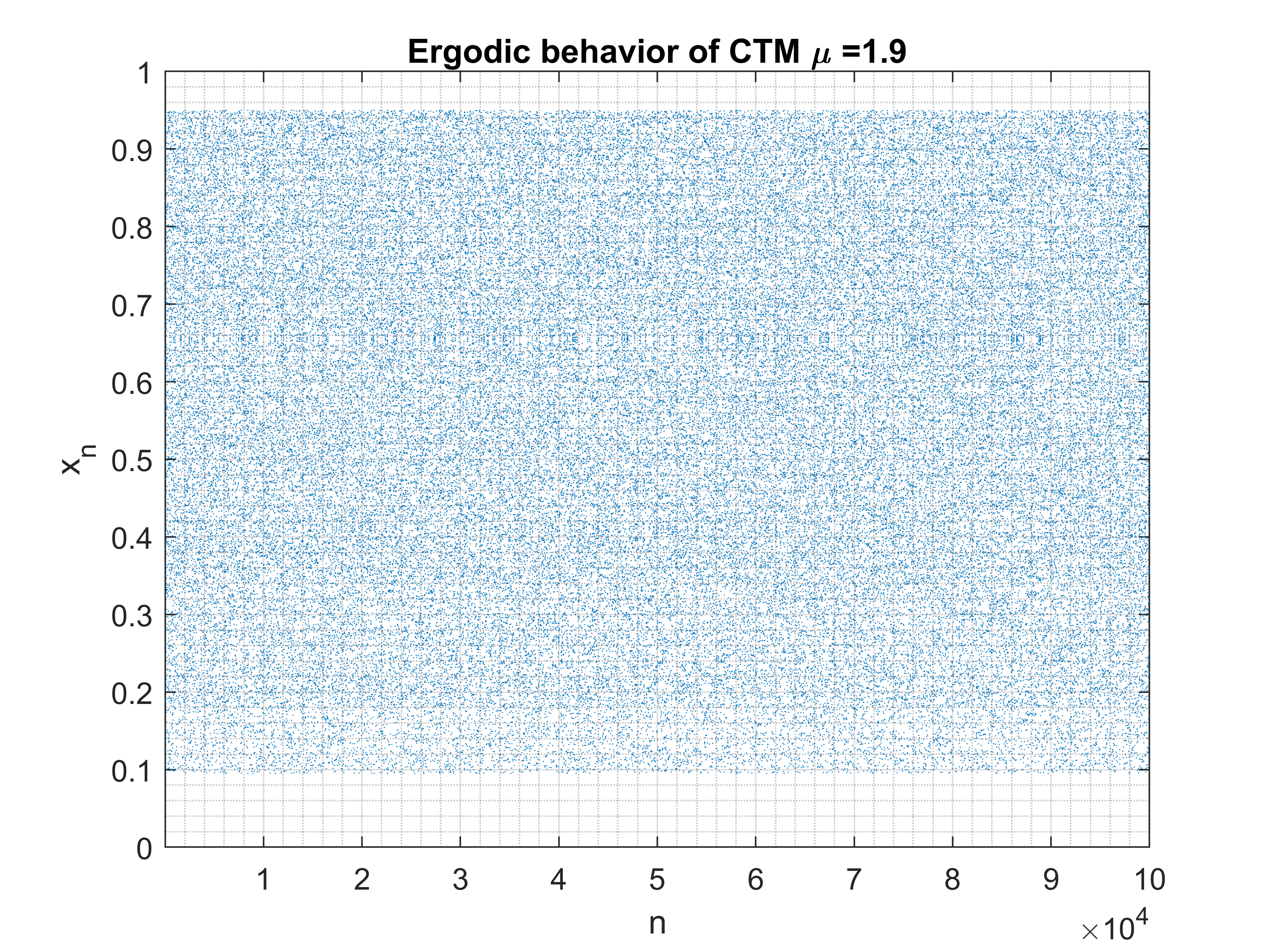}
    \caption{$\mu = 1.9$}
    \label{fig:image3}
  \end{subfigure}
  \hfill
  \begin{subfigure}{0.24\textwidth}
    \includegraphics[trim=0 0 0 22, clip, width=\linewidth]{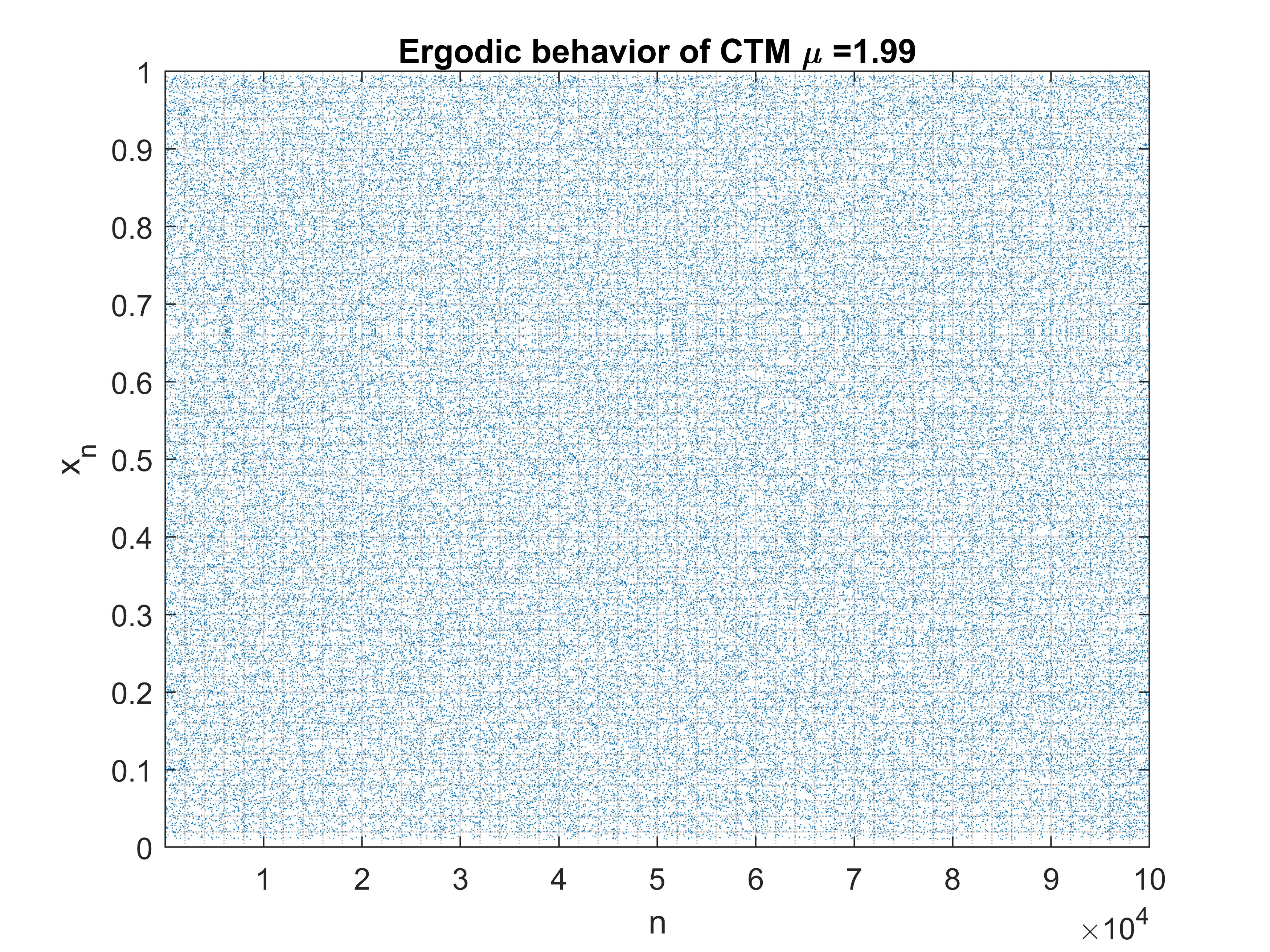}
    \caption{$\mu = 1.99$}
    \label{fig:image4}
  \end{subfigure}
  \caption{The plots illustrate the ergodic behavior of the CTM for different values of the control parameter $\mu$. The ergodicity is shown by the distribution of the values $x_n$ over iterations $n$. (a) $\mu = 1.2$ shows limited spread, indicating less ergodic behavior. (b) $\mu = 1.5$ demonstrates a wider spread. (c) $\mu = 1.9$ exhibits even more spread, and (d) $\mu = 1.99$ shows a near-uniform distribution, indicating highly ergodic behavior.}
  \label{fig:ergo_tent_map}
\end{figure*}

\begin{figure*}
  \centering
  \begin{subfigure}{0.24\textwidth}
    \includegraphics[trim=0 0 0 22, clip, width=\linewidth]{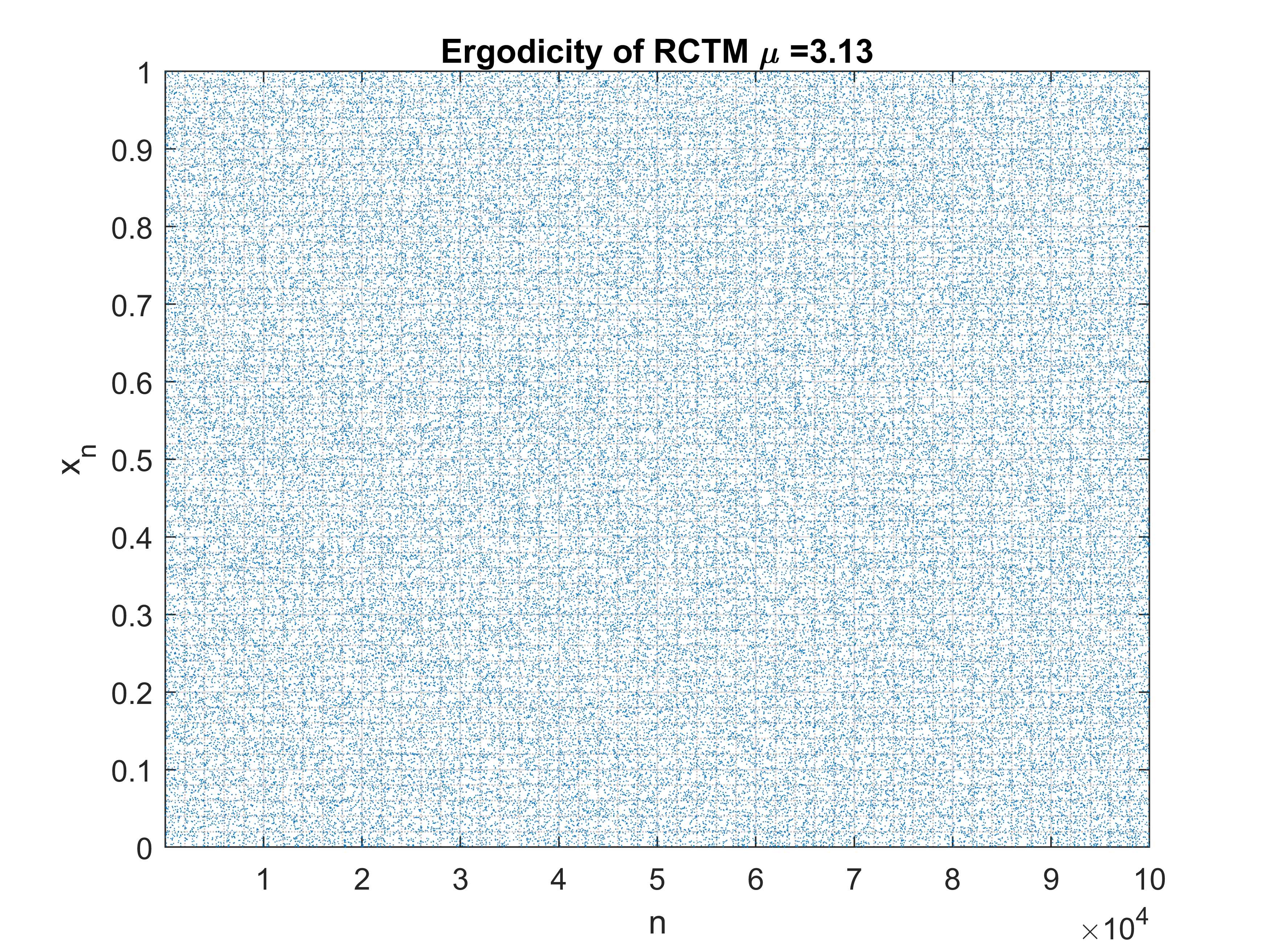}
    \caption{$\mu = 3.13$}
    \label{fig:subimage1}
  \end{subfigure}
  \hfill
  \begin{subfigure}{0.24\textwidth}
    \includegraphics[trim=0 0 0 22, clip, width=\linewidth]{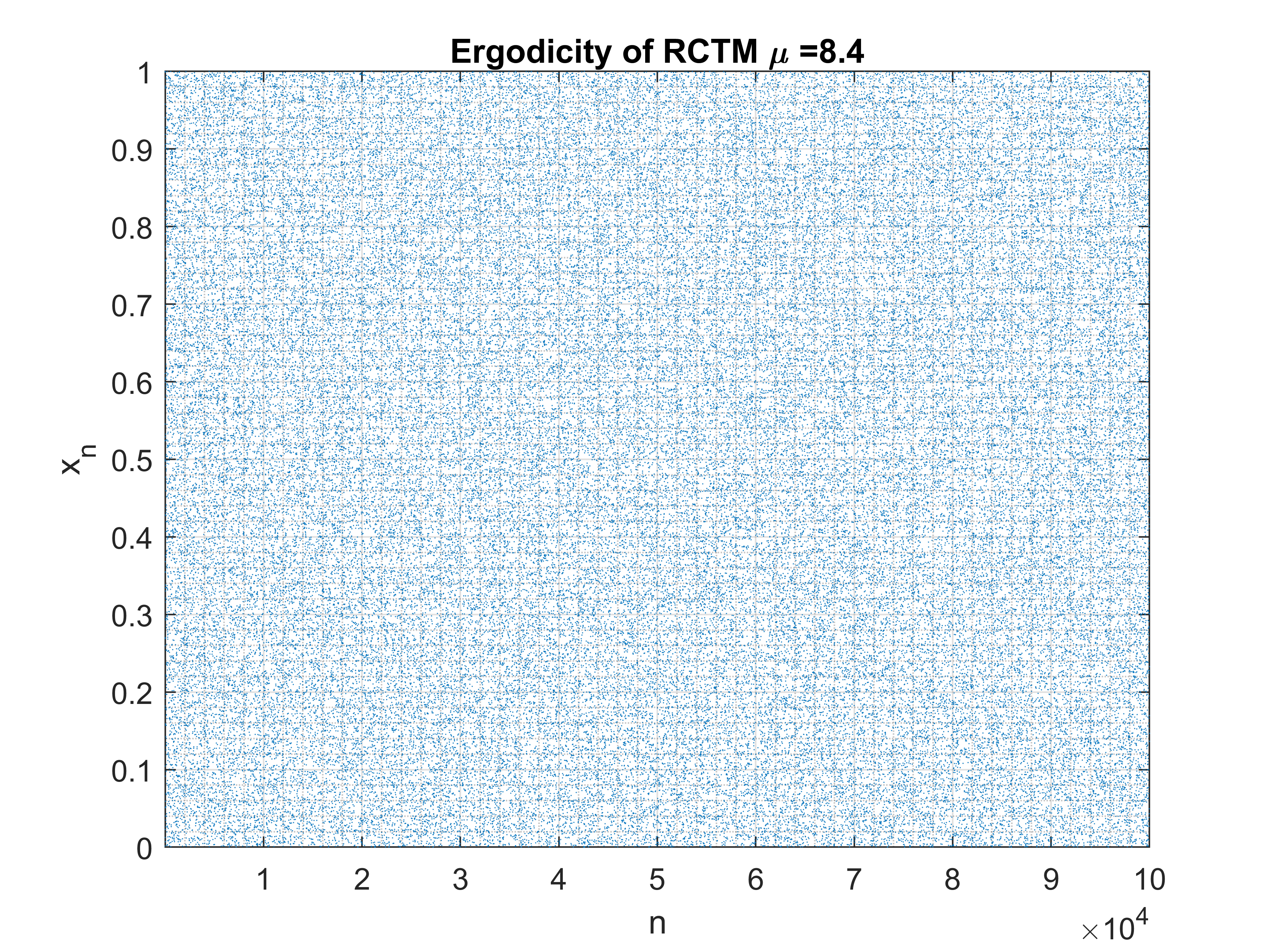}
    \caption{$\mu = 8.4$}
    \label{fig:image2}
  \end{subfigure}
  \hfill
  \begin{subfigure}{0.24\textwidth}
    \includegraphics[trim=0 0 0 22, clip, width=\linewidth]{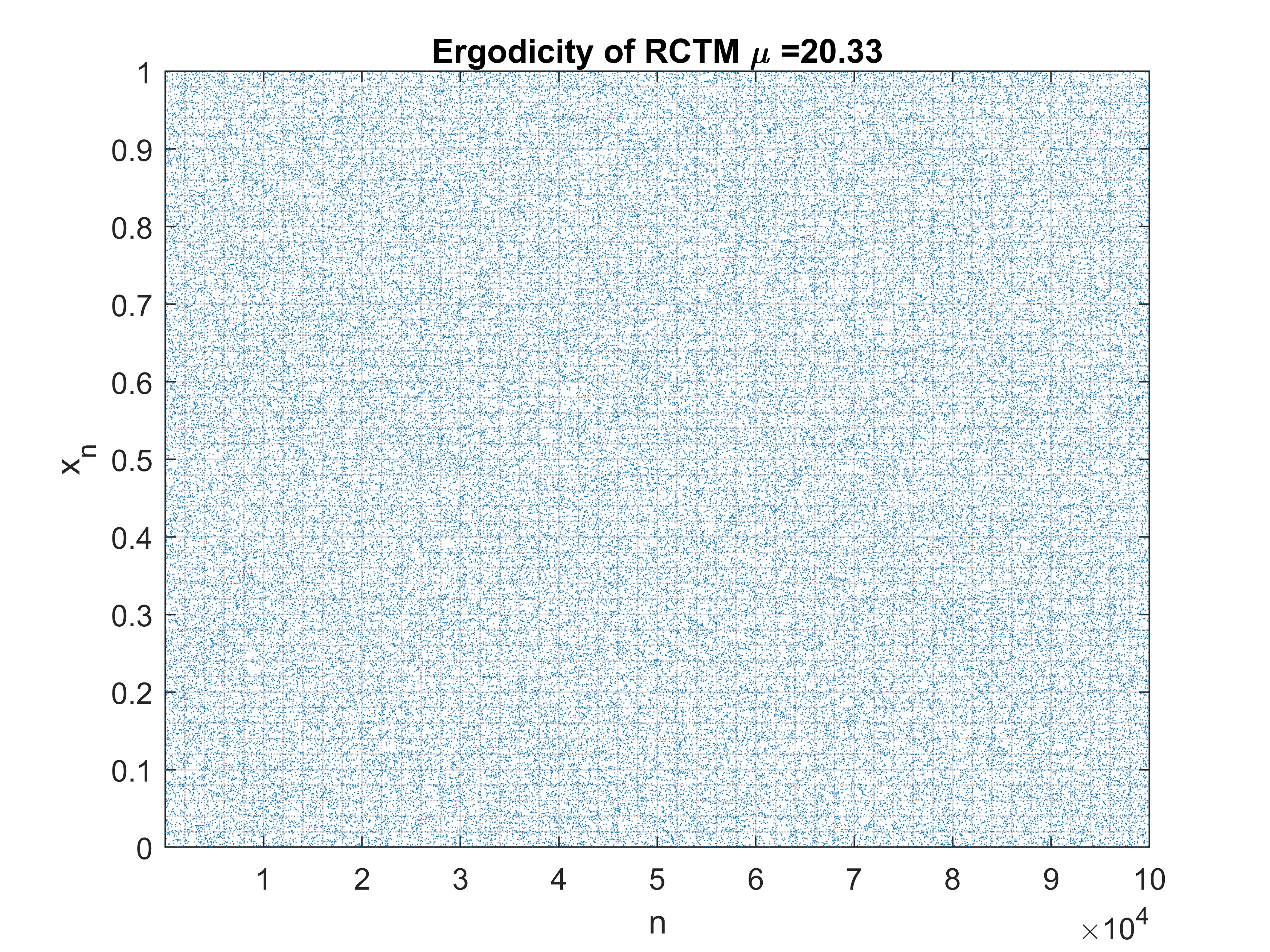}
    \caption{$\mu = 20.33$}
    \label{fig:image3}
  \end{subfigure}
  \hfill
  \begin{subfigure}{0.24\textwidth}
    \includegraphics[trim=0 0 0 22, clip, width=\linewidth]{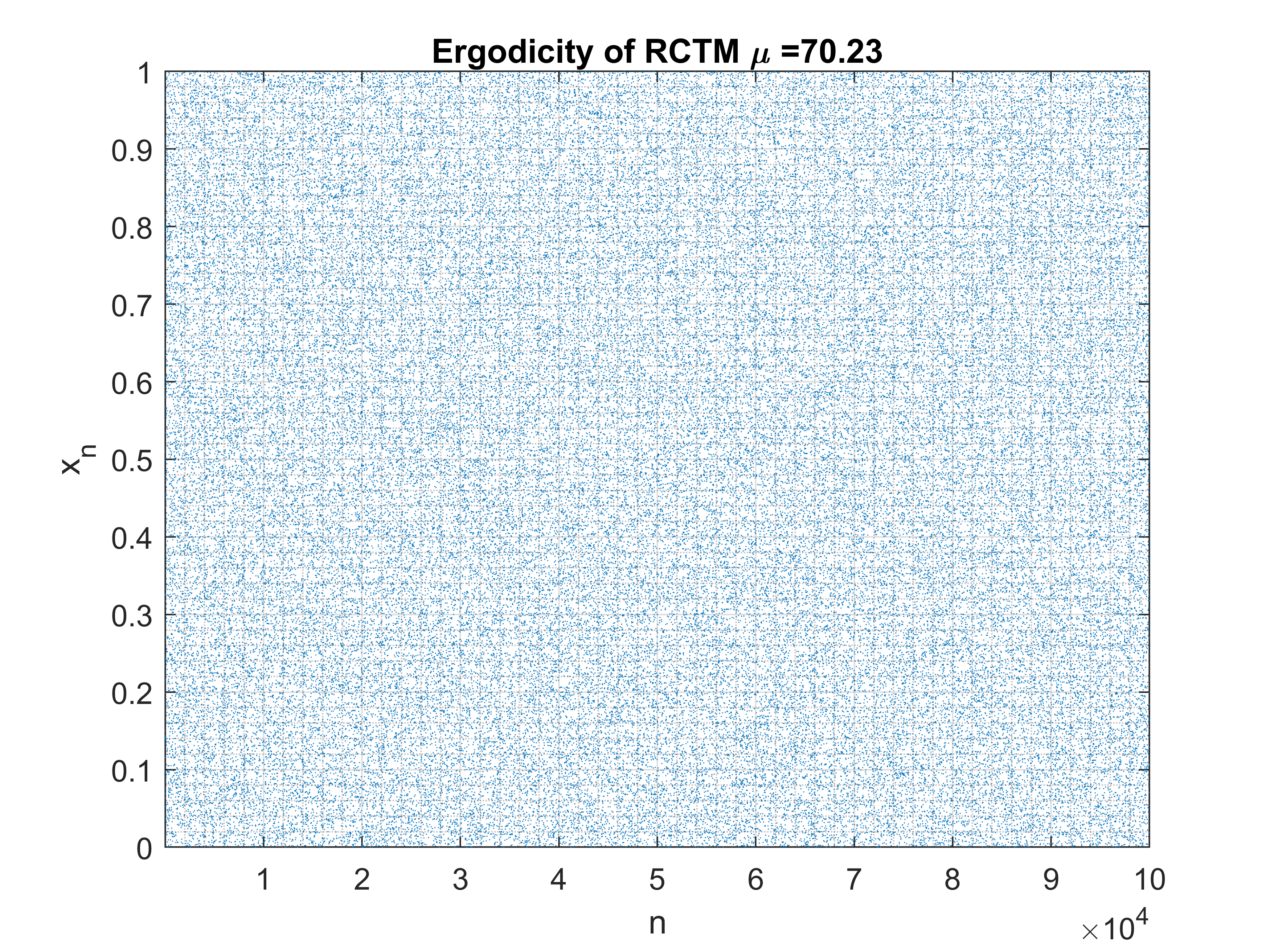}
    \caption{$\mu = 70.23$}
    \label{fig:image4}
  \end{subfigure}
  \caption{The plots illustrate the ergodic behavior of the RCTM for different values of the control parameter $\mu$. The ergodicity is shown by the distribution of the values $x_n$ over iterations $n$. (a) $\mu = 3.13$ shows a widespread distribution, indicating high ergodicity. (b) $\mu = 8.4$ maintains a similar ergodic spread. (c) $\mu = 20.33$ exhibits an extensive distribution, and (d) $\mu = 70.23$ shows a near-uniform distribution, highlighting the map's strong ergodic properties across various control parameter values.}
  
  \label{fig:ergodicity_modified_tent_map}
\end{figure*}

\subsection{Bifurcation Diagram}
\begin{figure}
    \centering
    \includegraphics[trim=0 0 0 22, clip, width=\columnwidth]{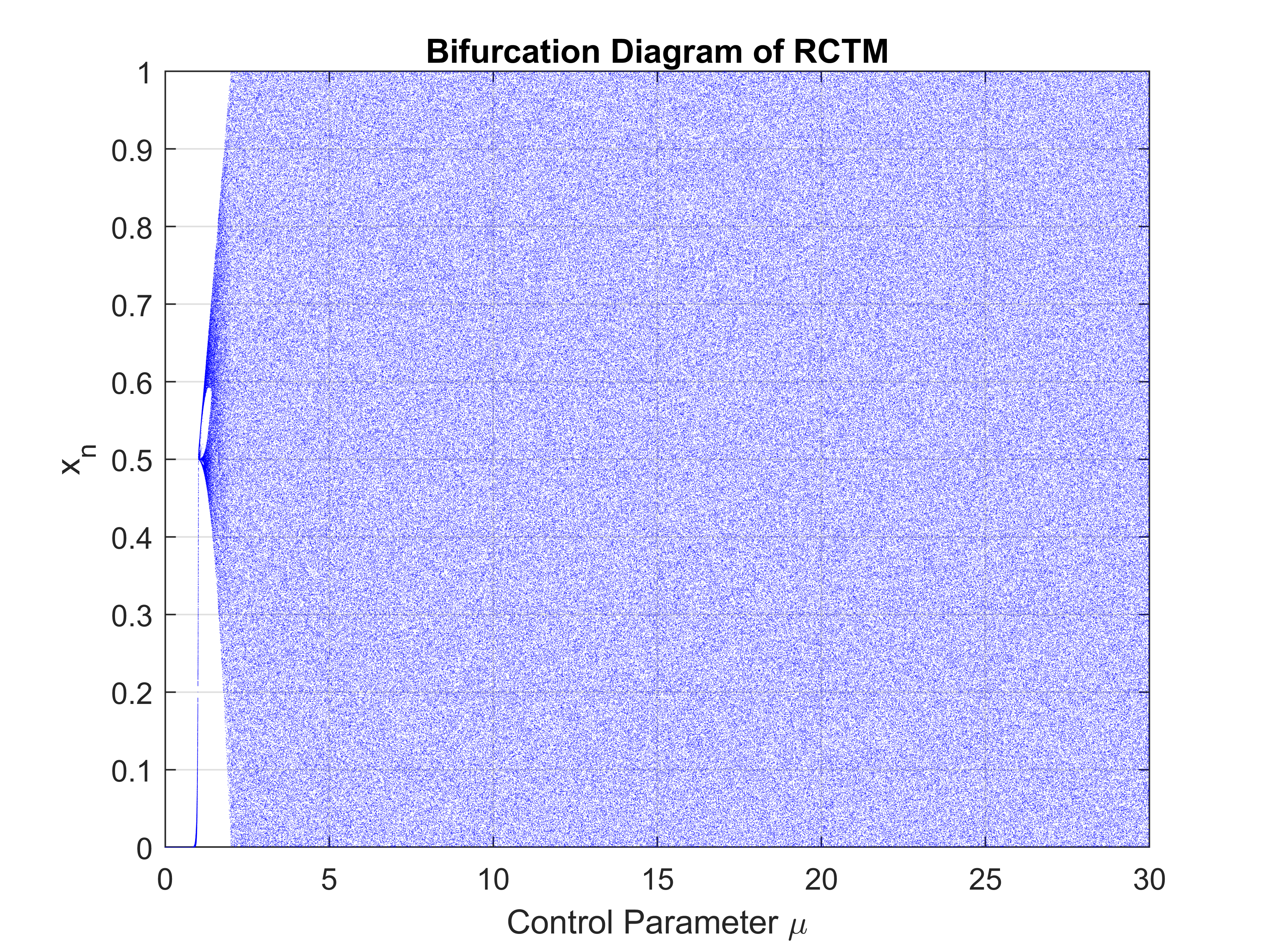}
    \caption{Bifurcation Diagram of the Robust Chaotic Tent Map (RCTM). This diagram illustrates the ergodic behavior of the RCTM by showing the distribution of the variable \( x_n \) over a range of control parameter values \( \mu \). As \( \mu \) increases, the system exhibits bifurcations leading to complex, chaotic behavior. The vertical axis represents the values of \( x_n \), while the horizontal axis represents the control parameter \( \mu \). This bifurcation diagram demonstrates the transition from ordered to chaotic states as \( \mu \) changes, highlighting the map's dynamic complexity and ergodic properties.}
    \label{fig:bifurcation_ctm}
\end{figure}

The bifurcation diagrams show the points on the vertical axis as the control parameter value is varied. The bifurcation of periodic orbits and chaotic attractors for CTM and \ac{RCTM} are demonstrated in Fig.~\ref{fig:tent_bifurcation} and Fig.~\ref{fig:bifurcation_ctm} respectively. The Fig.~\ref{fig:bifurcation_ctm} shows dense, chaotic attractor as the control parameter $\mu \in (0,100)$   is varied. As the $\mu > 2$ in CTM, the chaotic attractor become unstable and chaotic trajectories are vanished. Whereas, in \ac{RCTM}, new dense attractors exist in the neighborhood of parameter space that makes the behavior chaotic and occupies the entire phase space in the range (0,1).

\subsection{Lyapunov Exponent}

\begin{figure}
    \centering
    \includegraphics[trim=0 0 0 22, clip,width=\columnwidth]{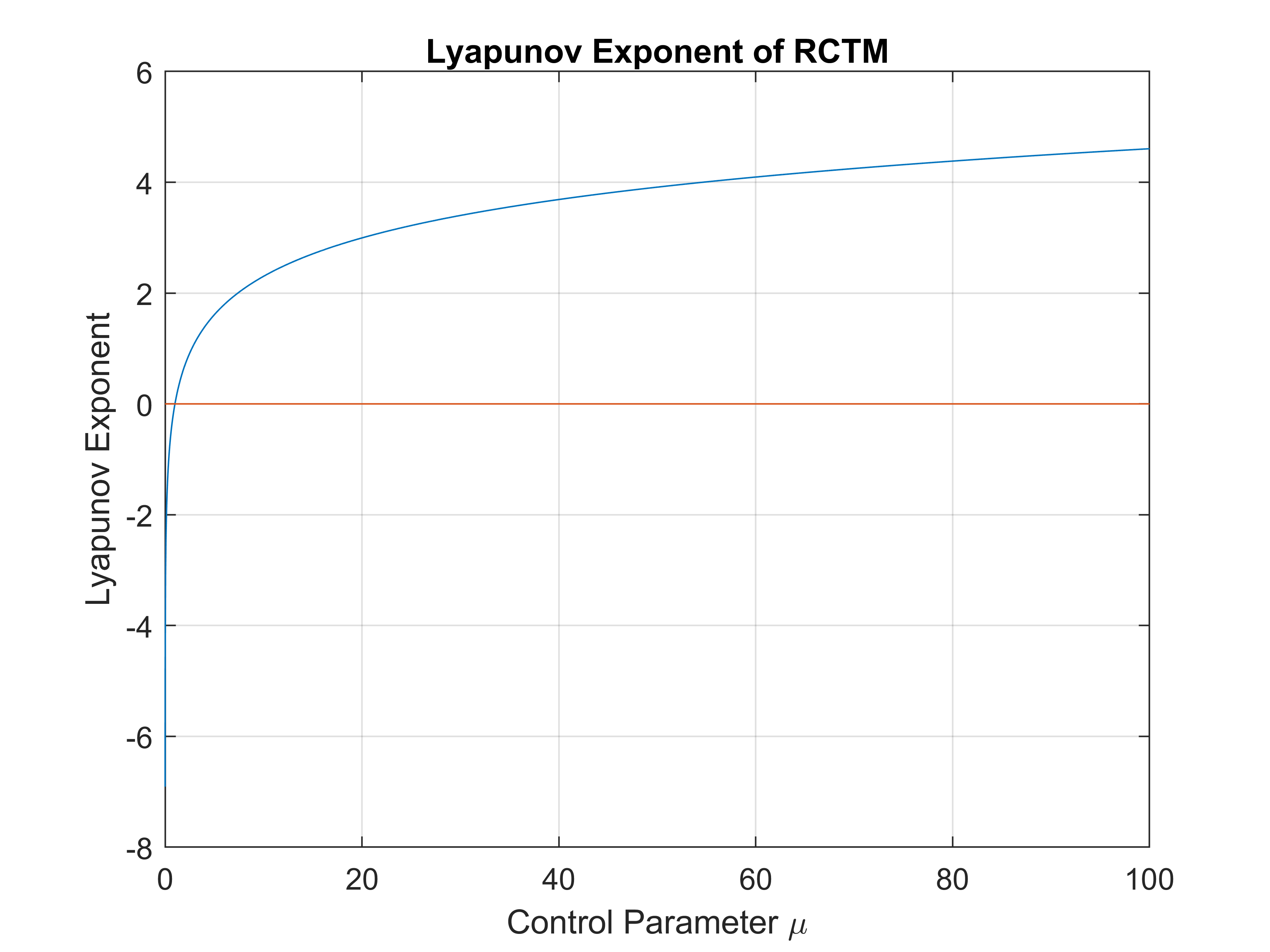}
    \caption{Lyapunov Exponent of the Robust Chaotic Tent Map (RCTM). This plot shows the Lyapunov exponent as a function of the control parameter \( \mu \). The Lyapunov exponent quantifies the sensitivity to initial conditions, with positive values indicating chaotic behavior. The vertical axis represents the Lyapunov exponent, while the horizontal axis represents the control parameter \( \mu \). As \( \mu \) increases, the Lyapunov exponent transitions from negative to positive, demonstrating the onset of chaos in the system. This figure highlights the dynamic complexity and ergodic properties of the \ac{RCTM} across a range of control parameter values.}
  
    \label{fig:laypnuv_ctm}
\end{figure}

The most exciting property of chaotic systems that makes them suitable for cryptography is sensitivity to control parameters and initial conditions. A highly nonlinear chaotic map generates extremely nonlinear trajectories, and a small perturbation in the initial condition or control parameter can generate extremely unpredictable trajectories. Thus, a positive value of the Lyapunov exponent indicates chaotic behavior and orbital divergence with a minimal change at the initial condition of the control parameter.  The Lyapunov exponent of \ac{RCTM} is shown in Fig.~\ref{fig:laypnuv_ctm} for $\mu \geq 1$. The Lyapunov exponent is positive for $\mu > 2$, which indicates that \ac{RCTM} has chaotic behavior.

\subsection{Ergodicity}

Another property of a chaotic map is mixing and ergodicity. It states that any chosen small set of initial points will eventually visit the entire phase space of the system when the map is iterated. It is observed in Fig.~\ref{fig:ergo_tent_map} that the CTM behavior is not ergodic for $\mu < 1.99$. The CTM shows ergodic behavior for $\mu =1.99$ with the phase space is fully covered. The Fig.~\ref{fig:ergodicity_modified_tent_map}  shows the ergodic behavior of RCTM. The RCTM has ergodic behavior shown in the figure, that verifies the coverage of complete phase space for  $\mu >2$.

\begin{figure*}
    \centering
    \includegraphics[trim= 0 40 10 10,clip,width=\textwidth]{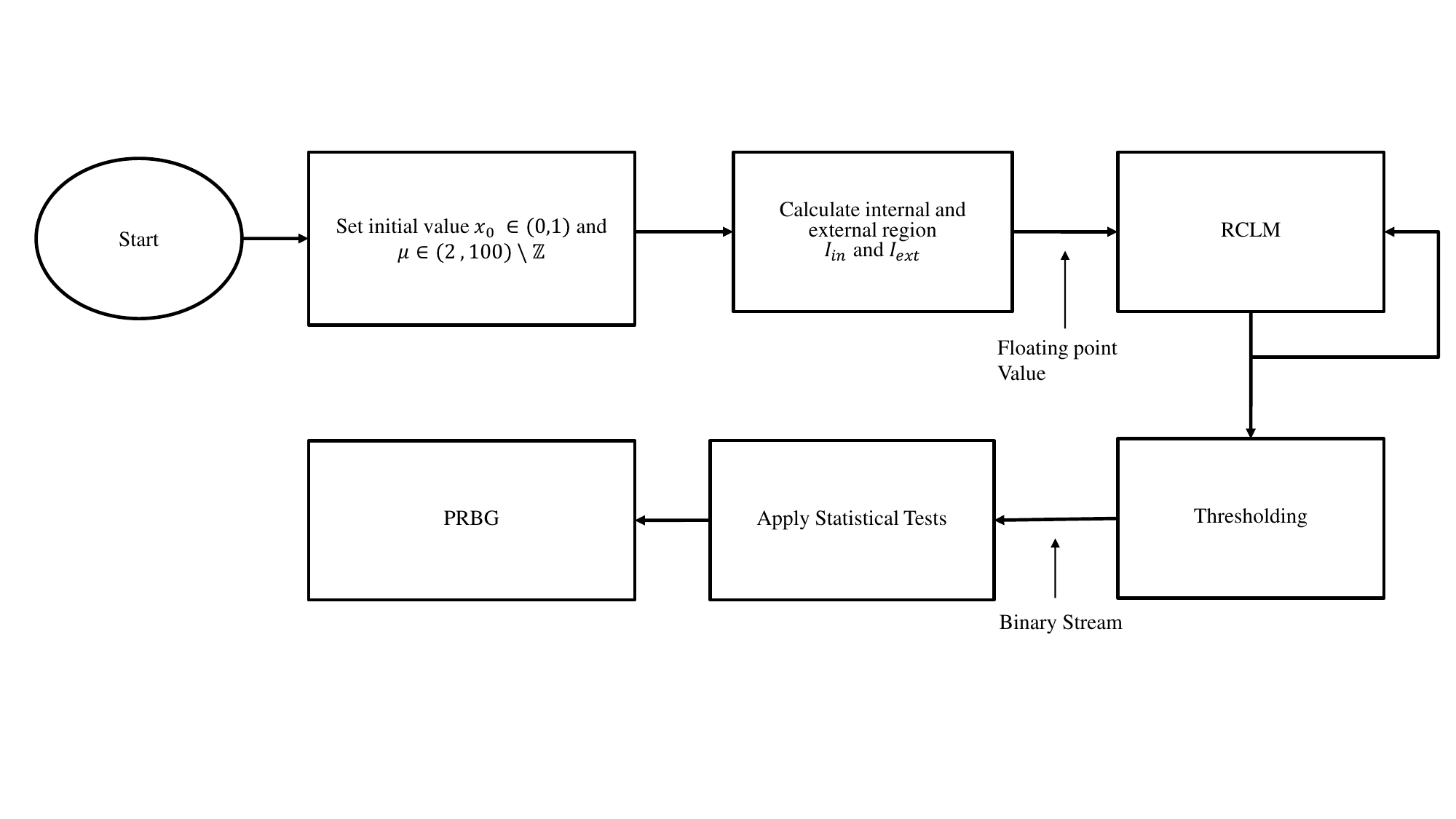}
    \caption{Proposed Methodology for Cryptographically Secure Pseudorandom Bit Generator (PRBG). The process starts with setting an initial value \( x_0 \in (0,1) \) and a control parameter \( \mu \in (2,100) \setminus \mathbb{Z} \). The internal and external regions \( I_{int} \) and \( I_{ext} \) are calculated, followed by the RCTM generating a sequence of floating-point values. These values are then subjected to a thresholding process to produce a binary stream. The resulting binary stream is used in the PRBG, which is then validated through various statistical tests to ensure its cryptographic security.}
  
    \label{fig:methodology}
\end{figure*}

% \section{Methodology: Pseudo-random Bit Generation Method}
% \label{sec:methodology}

% A Thresholding based simple approach is proposed using RCTM to design pseudo-random bit generation (PRBG). This proposed technique is presented in~\ref{fig:methodology}. and detailed steps are described below:
\begin{algorithm}
\caption{Algorithm for generating cryptographically secure pseudo-random number generation using Robust chaotic tent map}
\label{alg:tent_map}
\begin{algorithmic}[1]
\Procedure{robusttentmap}{$x_0, m, N$}
    \State $x \gets$ array of zeros with length $N$
    \State $x(1) \gets x_0$
    \State $n1 \gets 0.5 - \frac{\text{mod}(m/2,1)}{m}$
    \State $n2 \gets 0.5 + \frac{\text{mod}(m/2,1)}{m}$

    \For{$n \gets 1$ to $N-1$}
        \If{$m \leq 2$}
            \If{$x(n) < 0.5$}
                \State $x(n+1) \gets m \times x(n)$
            \Else
                \State $x(n+1) \gets m \times (1 - x(n)) $
            \EndIf
        \Else
            \If{$x(n) \geq n1$ \textbf{and} $x(n) \leq n2$}
                \If{$x(n) < 0.5$}
                    \State $x(n+1) \gets \frac{\text{mod}(m \times x(n) , 1)}{\text{mod}(m/2,1)}$
                \Else
                    \State $x(n+1) \gets \frac{\text{mod}(m \times (1 - x(n)) , 1)}{\text{mod}(m/2,1)}$
                \EndIf
            \Else
                \If{$x(n) < 0.5$}
                    \State $x(n+1) \gets \text{mod}(m \times x(n), 1)$
                \Else
                    \State $x(n+1) \gets \text{mod}(m \times (1 - x(n)) , 1)$
                \EndIf
            \EndIf
        \EndIf
    \EndFor

    \State \textbf{return} $x$
\EndProcedure
\end{algorithmic}
\end{algorithm}

% \textbf{Step 1}: Choose the initial condition $x_0 \in (0,1)$ and control parameter $\mu>2$ provided that $\mu \notin \mathbb{Z}$. These parameters are known as the secret keys of the proposed algorithm.

% \textbf{Step 2}: Iterate RCTM with chosen control parameter and arbitrary chosen initial condition, which is a secret key to generate a floating number of desired lengths sequence.

% \textbf{Step 3}: Apply threshold $\tau$ to generate the binary sequence, if $\tau <0.5$ clear bit for corresponding floating value and set bit if $\tau \geq 0.5$. 

% ~\ref{alg:generation_of_bitstream} and~\ref{alg:tent_map} explains above mentioned steps in the algorithmic form. Various statistical analyses, such as NIST STS, ENT, correlation, Net Pixel Change Rate (NCPR), etc., are applied to verify the proposed method's introduced randomness.  
% The following subsection discusses those performance analysis parameters in detail.

\begin{algorithm}
\caption{Algorithm for cryptographically secure pseudo-random bit generation}
\label{alg:generation_of_bitstream}
\begin{algorithmic}[1]
\State $m \in (2,100) $
\State $x_0 \in (0,1)$
\State $N$ bits required to generate

\State $x \gets$ \textsc{robusttentmap}($x_0, m, N$)

\State $outStream \gets$ \texttt{(x $\geq$ 0.5)} \Comment{Conversion to bitstream}

\end{algorithmic}
\end{algorithm}

\section{Methodology: Pseudo-random Bit Generation Method}
\label{sec:methodology}

This section outlines the methodology for designing a pseudo-random bit generator using \ac{RCTM}. The approach relies on thresholding and utilizes the inherent chaotic properties of the tent map to generate binary sequences. The process is visualized in Fig.~\ref{fig:methodology}.

\subsection{Algorithm Description}
The core of our methodology is encapsulated in the~\ref{alg:generation_of_bitstream} and \ref{alg:tent_map} for \ac{CSPRNG}, detailed below:

\subsubsection{Initialization}
Choose an initial condition $x_0$ within the open interval (0,1), and a non-integer control parameter $\mu$ greater than 2. These parameters serve as the secret keys for the algorithm, ensuring the uniqueness and unpredictability of the generated sequence.

\subsubsection{Tent Map Iteration}
Apply the \ac{RCTM} iteratively using the chosen control parameter and initial condition. This iteration produces a sequence of floating-point numbers, each representing a state of the system at a particular step.

\subsubsection{Threshold Application}
Convert the sequence of floating-point values into a binary stream. This conversion uses a threshold $\tau = 0.5$; values below the threshold are assigned a bit of 0, and values above or equal to the threshold are assigned a bit of 1.

% The detailed steps of the algorithm are formally presented in~\ref{alg:tent_map}, and the bit generation process in~\ref{alg:generation_of_bitstream}.

\subsection{Statistical Validation}
To ensure the randomness of the generated binary sequences, various statistical tests are applied, including the NIST Statistical Test Suite (STS), ENT, and measures of correlation and Net Pixel Change Rate (NCPR). The performance analysis and results of these tests are discussed in the subsequent sections.

This methodological framework not only provides a robust mechanism for generating pseudo-random bits but also ensures cryptographic security through its dependency on chaotic dynamics and thresholding.

\section{Statistical Analysis of proposed CS-PRNG using RCTM}\label{sec:analysis}

\subsection{NIST Statistical test suite Analysis}
Currently, NIST STS is used to find the randomness of the produced bitstream. This test suite is comprised of 15 statistical tests, and each test returns a p-value. If the p-value for a test is greater than the specified significant value, and then the test is considered random. A bitstream must pass all 15 tests for the cryptographic application. Here, we have produced 40 bitstreams of each $10^6$ bits length at $\mu=61.81$ and $x_0=0.23$. The generated bitstream passes all the tests with a significant p-value; the test results are shown in Table~\ref{tab:nist_results}.

\begin{table*}
    \centering
    \caption{NIST Statistical Test Results for RCTM at $\mu = 61.81$ and $x_0 = 0.23$. Each test includes its p-value, status (Passed/Failed), and the proportion of successful runs out of 20. The results demonstrate that RCTM consistently passes all NIST tests, indicating strong randomness properties. Note: Some tests provide results for both forward and reverse sequences. The asterisk (*) represents the average value of the p-values for the corresponding test.}

    %\resizebox{\textwidth}{!}{%
    \begin{tabular}{lccc}
    \hline
        \textbf{NIST Statistical Test} & \textbf{P-value} & \textbf{Status} & \textbf{Proportion } \\ \hline
        \textbf{Frequency (mono bit)} & 0.652785 & Passed & 20/20  \\ 
        \textbf{Block Frequency} & 0.408615 & Passed & 20/20  \\ 
        \textbf{Cumulative Sum} & 0.325117 (Forward) & Passed & 20/20  \\ 
        \textbf{} & 0.678729 (Reverse) & ~ &   \\ 
        \textbf{Longest Run} & 0.746859 & Passed & 20/20  \\ 
        \textbf{Runs} & 0.240045 & Passed & 20/20  \\ 
        \textbf{Rank} & 0.081801 & Passed & 20/20  \\ 
        \textbf{Non-overlapping Template Matchings} & 0.48082 * & Passed & 20/20  \\ 
        \textbf{Discrete Fourier Transform} & 0.163536 & Passed & 20/20  \\ 
        \textbf{Overlapping Template Matchings} & 0.015876 & Passed & 20/20  \\ 
        \textbf{Universal Statistical} & 0.201292 & Passed & 20/20  \\ 
        \textbf{Serial} & 0.087536 & Passed & 20/20  \\ 
        \textbf{} & 0.313509 & ~ &   \\ 
        \textbf{Random Excursions Variant} & 0.596007 * & Passed & 20/20  \\ 
        \textbf{Random Excursions} & 0.394827889 * & Passed & 20/20  \\ 
        \textbf{Approximate Entropy} & 0.531342 & Passed & 20/20  \\ 
        \textbf{Linear Complexity} & 0.685106 & Passed & 20/20  \\ 
        \hline
    \end{tabular}
    %}
    \label{tab:nist_results}
\end{table*}

\subsection{ENT Statistical Test Suite}
ENT statistical test suit is designed to predict the randomness of sequence, comprising six tests~\cite{Walker}. To support our proposed methodology with ENT, we have generated a byte stream of 10 million. We applied the same control parameters as we applied to the NIST test suite. Table~\ref{tab:ent_results} shows that the successful results for ENT that are achieved using the proposed scheme.

\begin{table}
    \centering
    \caption{ENT Statistical Test Results of RCTM. All results indicate strong randomness: Entropy is close to 8 for one byte, optimum compression is at 0\%, Chi-Square results fall within the 10\%-90\% range, the arithmetic mean is close to 127.5, the Monte Carlo value for $\pi$ shows minimal error, and the serial correlation coefficient is near zero.}
    
    \begin{tabular}{lcc}
    \hline
        \textbf{Statistical Test} & \textbf{Conditions} & \textbf{Results} \\ \hline
        \textbf{Entropy} & $\approx 8$ (for one byte) & $7.999$ \\ 
        \textbf{Optimum Compression} & $\approx 0\%$ & 0\% \\ 
        \textbf{Chi-Square} & 10\%-90\% & 20.24\% \\ 
        \textbf{Arithmetic Mean} & 127.5 & 127.4502 \\ 
        \textbf{Monte Carlo Value for $\pi$} & $\approx 0\%$ error & 0.03\% \\ 
        \textbf{Serial Correlation Coefficient} & $\approx 0$ & -0.000038 \\ 
        \hline
    \end{tabular}
    
    \label{tab:ent_results}
\end{table}

\subsection{TestU01 Test Suite}
TestU01~\cite{l2007testu01} is written in C and designed to determine the random behavior of \ac{PRNG}s. We adopted three test batteries, \textit{Rabbit}, \textit{Alphabit}, and \textit{BlockAlphabit}, to measure the randomness of binary sequences generated by our proposed methodology. Table~\ref{tab:testu01_result} presents the results of TestU01 on two floating-point sequences of lengths \(2^{20}\) and \(2^{30}\). Based on the results shown in Table~\ref{tab:testu01_result}, we can claim that the sequences generated by our study can be used for pseudo-random number generation.

\begin{table}
   \centering
   \caption{Results of the TestU01 test suite. The tests were performed on sequences of sizes \(2^{20}\) and \(2^{30}\). Partial tests are indicated by an asterisk (*).}
   \resizebox{\columnwidth}{!}{%
     \begin{tabular}{cccll}
     \hline
     \multicolumn{1}{l}{\textbf{Battery}} & \multicolumn{1}{l}{\textbf{Number of Tests}} & \multicolumn{1}{l}{\textbf{Parameters}} & \textbf{Sequence Size} & \textbf{Result} \\
     \hline
     \multirow{2}[1]{*}{Rabbit} & \multirow{2}[1]{*}{28} & \multirow{2}[1]{*}{Standard} & $2^{20}$ & Pass \\
           &       &       & $2^{30}$*  & Pass \\
     \multirow{2}[0]{*}{Alphabit} & \multirow{2}[0]{*}{17} & \multirow{2}[0]{*}{Standard} & $2^{20}$ & Pass \\
           &       &       & $2^{30}$*  & Pass \\
     \multirow{2}[1]{*}{Block Alphabit} & \multirow{2}[1]{*}{17} & \multirow{2}[1]{*}{Standard} & $2^{20}$ & Pass \\
           &       &       & $2^{30}$* & Pass \\
     \hline
     \end{tabular}%
     }
   \label{tab:testu01_result}%
 \end{table}%

\subsection{Key Space Analysis}

A strong cryptographic primitive must have $2^{100}$ key space to resist the exhaustive key search attack~\cite{alvarez2006some}. Our proposed study is comprised of two independent parameters and two dependent control parameters. The independent control parameters $x_0  \in (0,1)$, $\mu \in (2,100) \setminus \mathbb{Z}$, and dependent control parameters $n_1 \in \left( \frac{1}{2} - \left( \mu mod 1 \right) / \left(\mu \right),1/2  \right)$ and $n_2\in(\frac{1}{2}  ,\frac{1}{2}+(\mu mod 1)/(\mu ) )$. The IEEE standard stated in~\cite{kahan1996ieee}  double floating-point precision is $10^{-16}$.  Therefore, key space for $x_0= 10^{16}$, $\mu =(100-2)\times10^{16}  \simeq 10^{18}$ ,  $n_1 \simeq 10^{13}$ and $n_2\simeq 10^{13}$. Thus, the collective key space of the proposed technique is $ \simeq10^{16} \times 10^{18} \times 10^{13} \times 10^{13} \simeq 10^{60}= 2^{199}$. There exist some weak keys as well, if we consider half of the keys are weak keys, the acceptable keys are $2^{198}$ which is still meets the requirements of minimum key-space to resist the brute force attack. Table~\ref{tab:key_compare} shows the comparison of key-space with existing systems. Although some state of the art work has high keys but they depend on multiple parameter which has higher number of keys as well as disadvantages of managing such parameters.

\begin{table}
    \centering
    \caption{Comparison of Key Space and Control Parameters for Various Pseudorandom Number Generation Techniques. CP denotes the number of control parameters used in each technique.}
    \resizebox{\columnwidth}{!}{%
    \begin{tabular}[\columnwidth]{lcc}
    \hline
        \textbf{Existing Techniques} & \textbf{Key Space } & \textbf{Number of CP} \\ \hline
        Proposed (Ours) & $2^{198}$  & 2 \\ 
        Irfan et al.~\cite{irfan2020pseudorandom} & $2^{111}$  & 2\\ 
        Wang et al.~\cite{wang2016pseudorandom} & $2^{135}$   & 4 \\ 
        Murrilo-Escobar et al.~\cite{murillo2017novel} & $2^{128}$ & 2  \\ 
        Behnia et al.~\cite{behnia2011novel} & $2^{152}$   & 3 \\ 
        Wang et al.~\cite{wang2019pseudo} & $2^{160}$  & 3 \\ 
        Zia et al.~\cite{zia2023resource} & $2^{266}$  & 5 \\ 
        Zia et al.~\cite{zia2023resource} & $2^{372}$  & 7 \\ 
        \hline
    \end{tabular}
    }
    
    \label{tab:key_compare}
\end{table}

\subsection{Correlation Analysis}

The correlation coefficient is used to find the relationship between the two generated sequences. The generated sequences are compared to each other with the following Eq.~\eqref{eq:correaltion_coeff}:

\begin{equation}
r_{xy} = \frac{ L \sum_{i=0}^L \left(x_iy_i\right) - \sum_{i=0}^L x_i \sum_{i=0}^L y_i}{ \sqrt{ \left( L \sum_{i=0}^L x_i^2 - \left( \sum_{i=0}^L x_i\right)^2  \right) \left( L \sum_{i=0}^L y_i^2 - \left( \sum_{i=0}^L y_i\right)^2  \right) }} 
\label{eq:correaltion_coeff}
\end{equation}

\begin{figure}
    \centering
    \includegraphics[width=\columnwidth]{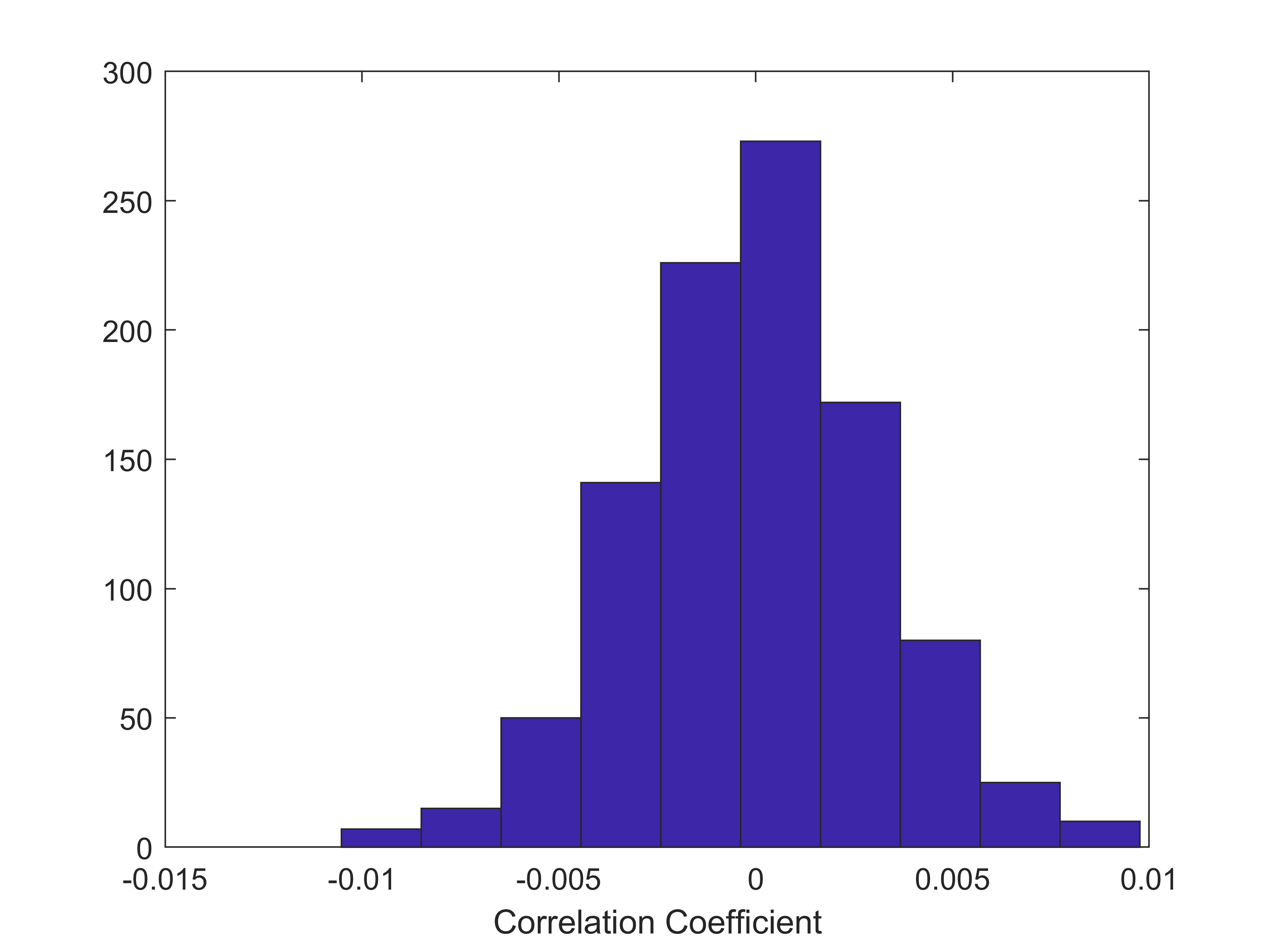}
    \caption{Correlation Coefficient Distribution}
    \label{fig:correlation_distribution}
\end{figure}

\begin{figure}
    \centering
    \includegraphics[width=\columnwidth]{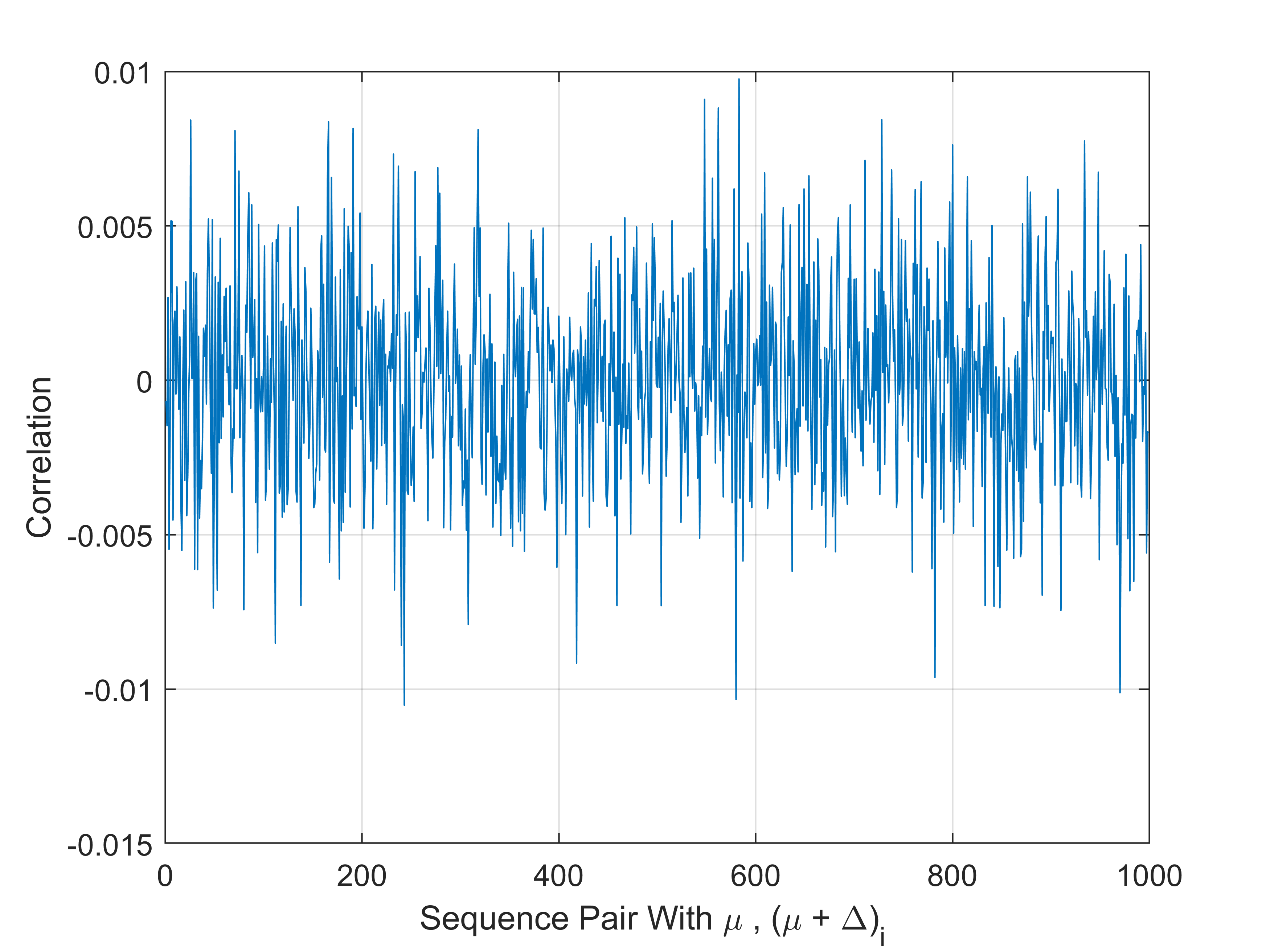}
   \caption{Correlation Analysis of RCTM Sequence Pairs. This plot shows the correlation between pairs of sequences generated by the Robust Chaotic Tent Map (RCTM) with control parameters $\mu$ and $(\mu + \Delta)$. The horizontal axis represents the sequence pairs, and the vertical axis represents the correlation values. The near-zero correlation values indicate that the sequences are uncorrelated, demonstrating the effectiveness of the RCTM in generating pseudorandom sequences.}
  
    \label{fig:correlation}
\end{figure}

Here $x_i$ and $y_i$ are two generated sequences of identical length L. The correlation $r_{xy} \in (-1,1)$, closer the value to zero of correlation coefficient makes it statistically independent from each other. We have generated $10^5$ sequences; each sequence comprises $10^3$ samples by adding a $\Delta = \frac{1}{2^{48}}$ to the initial conditions $\mu=61.81$ and $x_0=0.23$. The minimum and maximum correlation coefficients are $-0.0105$ and $0.0098$, respectively, as shown in~\ref{tab:correlation}. The uniform results of the correlation coefficient close to zero can be observed in Fig.~\ref{fig:correlation_distribution} and Fig.~\ref{fig:correlation}. 

\begin{table}[!ht]
    \centering
     \caption{Correlation Coefficient between Generated Sequences. The table shows the minimum and maximum correlation coefficients for two different cases, indicating the degree of linear relationship between the sequences. Lower correlation values suggest higher randomness.}
    
    \resizebox{\columnwidth}{!}{%
    \begin{tabular}{lcc}
    \hline
        \textbf{Correlation} & \textbf{Case 1} & \textbf{Case 2 } \\ \hline
        \textbf{Minimum Correlation} & -0.0043 & -0.0458  \\ 
        \textbf{Maximum Correlation} & 0.0245 & -0.0050  \\ \hline
    \end{tabular}
    }
    \label{tab:correlation}
\end{table}

\subsection{Information Entropy Analysis}

A proposed PRBG must produce a highly unpredictable binary sequence with high nonlinearity. The entropy is used to measure the amount of unpredictability in a given sequence. If an N-bit sequence is generated, it must hold a $2^N$ possible symbols. In an ideal case, the amount of information entropy should be equal to N. Following Eq.~\ref{eq:info_entroy} is used to find the entropy of a sequence $x_i$ with the probability of $P\left( x_i \right)$;
\begin{equation}
    H(X) = \sum_{i=0}^{2^N-1} P\left( x_i \right) \times log_2\left( \frac{1}{P\left( x_i\right)}\right)
    \label{eq:info_entroy}
\end{equation}

\begin{figure}
    \centering
    \includegraphics[width=\columnwidth]{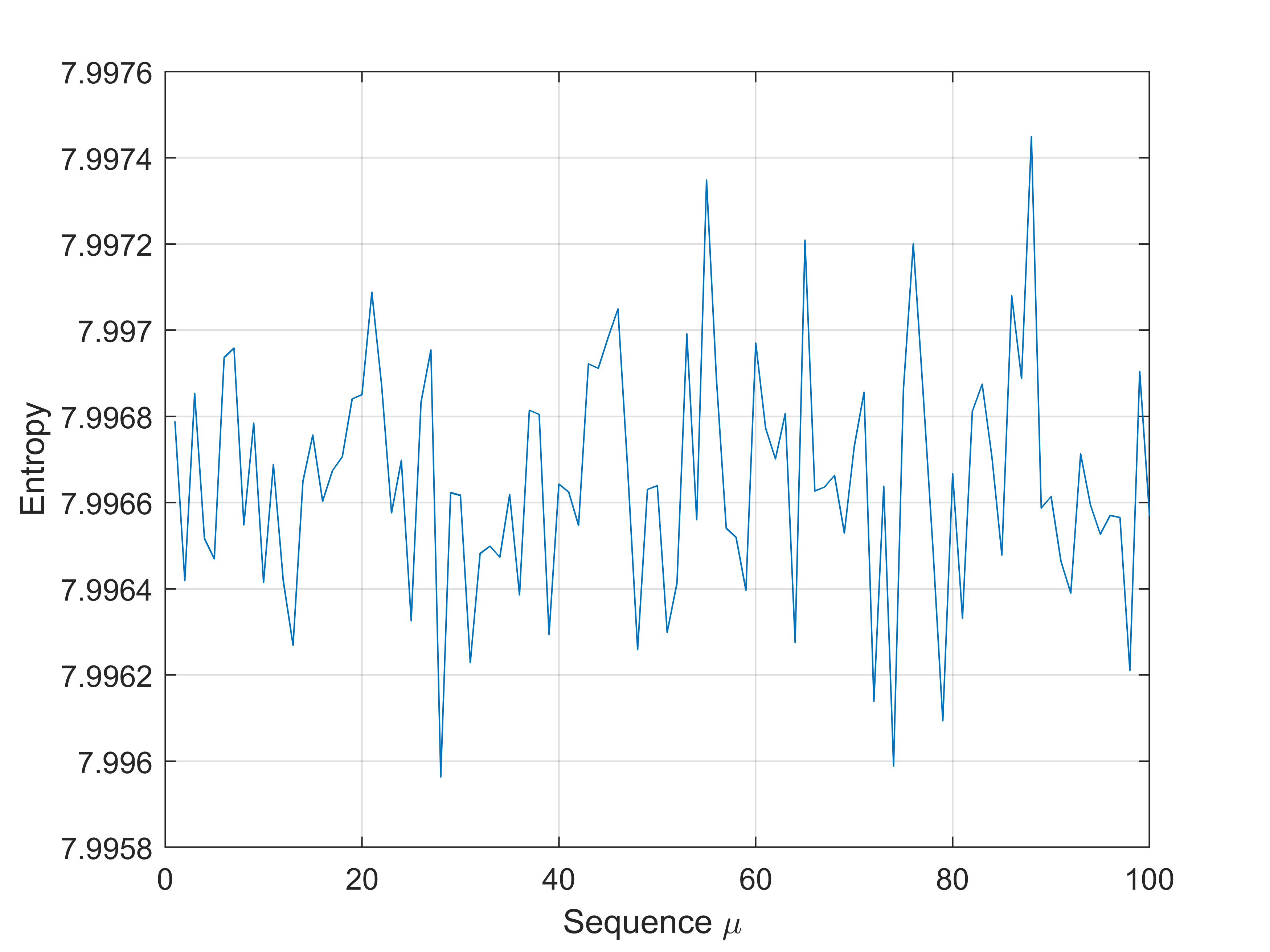}
    \caption{Information Entropy Analysis}
    \label{fig:entropy}
\end{figure}

To witness the randomness of the proposed method, we have produced 100 sequences, and each sequence consists of $10^5$ samples. Then the generated sequence is converted to 8-bit of each sample element, as shown in Fig.~\ref{fig:entropy} and the average entropy of the generated sequence is 7.9967. Therefore, it is another evidence that the proposed methodology generates a highly pseudo-random sequence. 

\subsection{Histogram Analysis}

\begin{figure}
    \centering
    \includegraphics[width=\columnwidth]{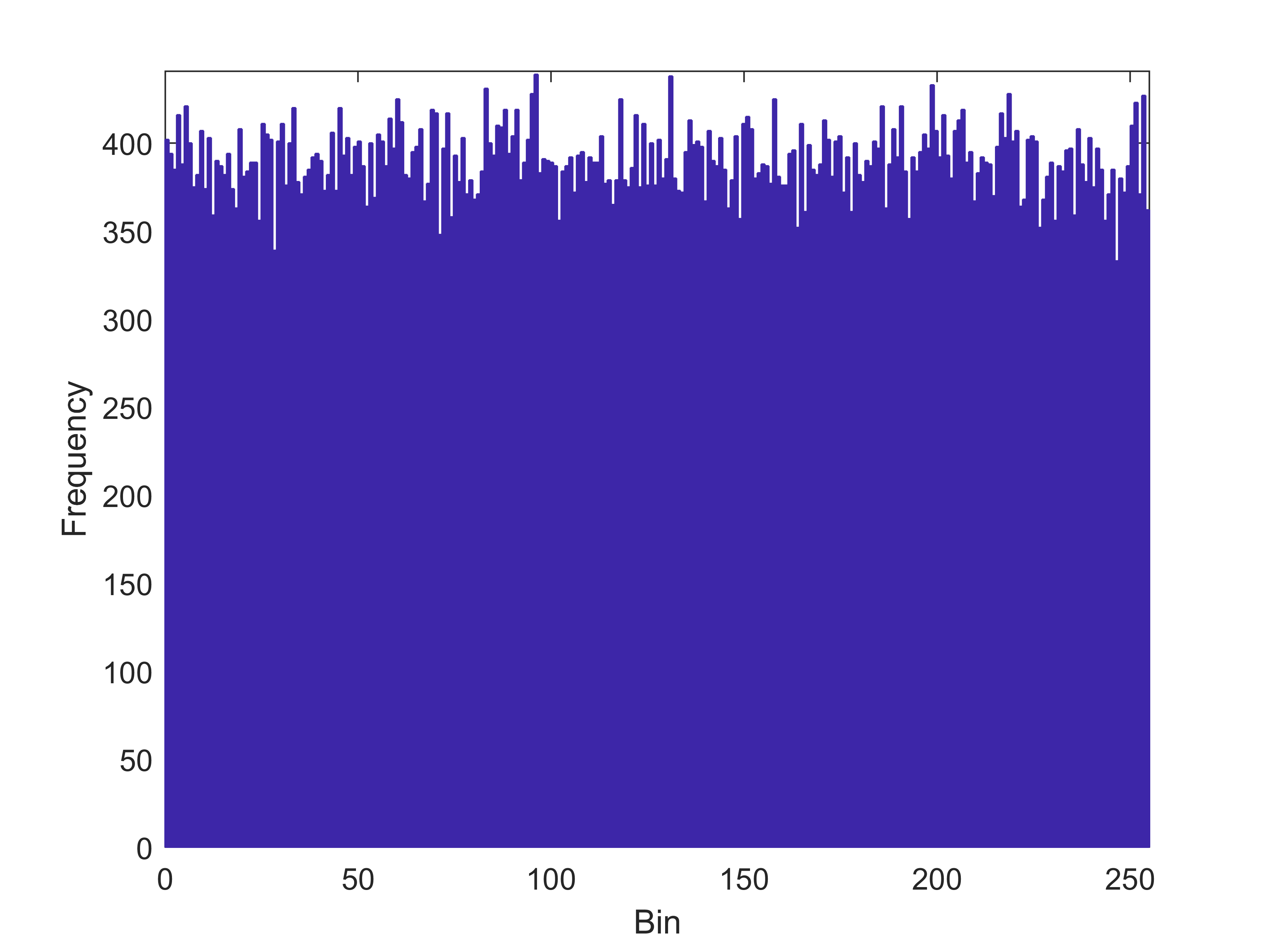}
    \caption{Histogram of the randomly generated sequence}
    \label{fig:histogram}
\end{figure}

A histogram shows the frequency of elements within a sequence in graphical form. A pseudo-random sequence must produce the even distribution of each sample which favors cryptographic primitives. In Fig.~\ref{fig:histogram}, the histogram is generated for $10^5$ sequences which shows that the uniform output. Hence, the proposed method produces all phase space elements with uniform distribution.

\subsection{Key Sensitivity Analysis}

Key sensitivity is the measure of small perturbation in the input that causes a substantial change to the output. For key sensitivity analysis, a similar procedure is opted in~\cite{wang2016pseudorandom}. For key sensitivity analysis of the proposed PRBG, we considered cases. In each case, prescribed parameters are selected. The chosen parameter setting for each case is described as follows:
\textbf{Case 1}: $\mu$ is iterated for $\mu= 49.13$ to $\mu+\Delta$ with $x_0=0.28$
\textbf{Case 2}: $x_0$ is repeated for $x_0= 0.28$ to $x_0+\Delta$ with $\mu=49.13$

\begin{figure}
  \centering
  \begin{subfigure}{0.45\textwidth}
    \includegraphics[width=\linewidth]{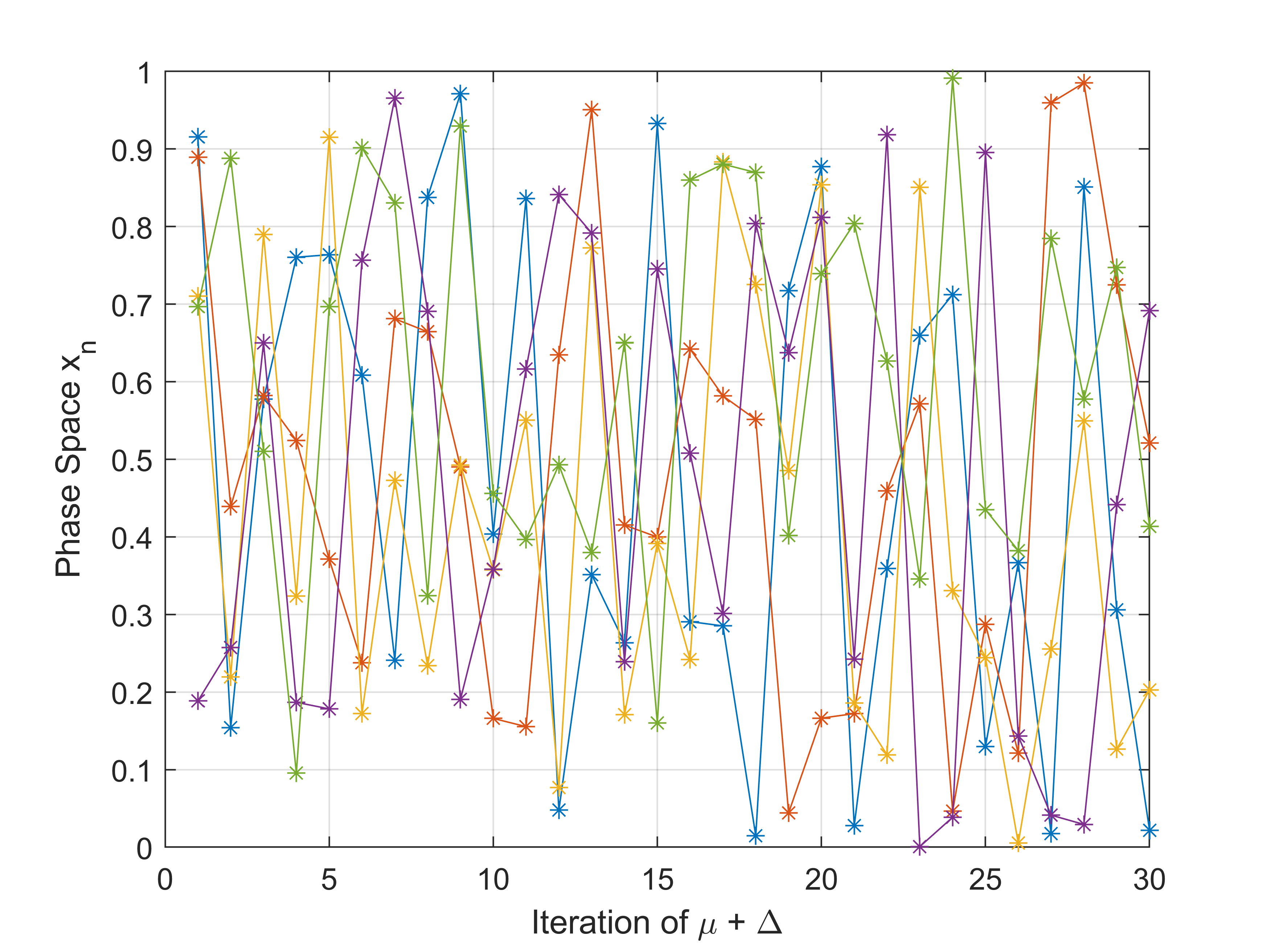}
    \caption{}
    \label{fig:figure_a}
  \end{subfigure}
  \hspace{0.05\textwidth}
  \begin{subfigure}{0.45\textwidth}
    \includegraphics[width=\linewidth]{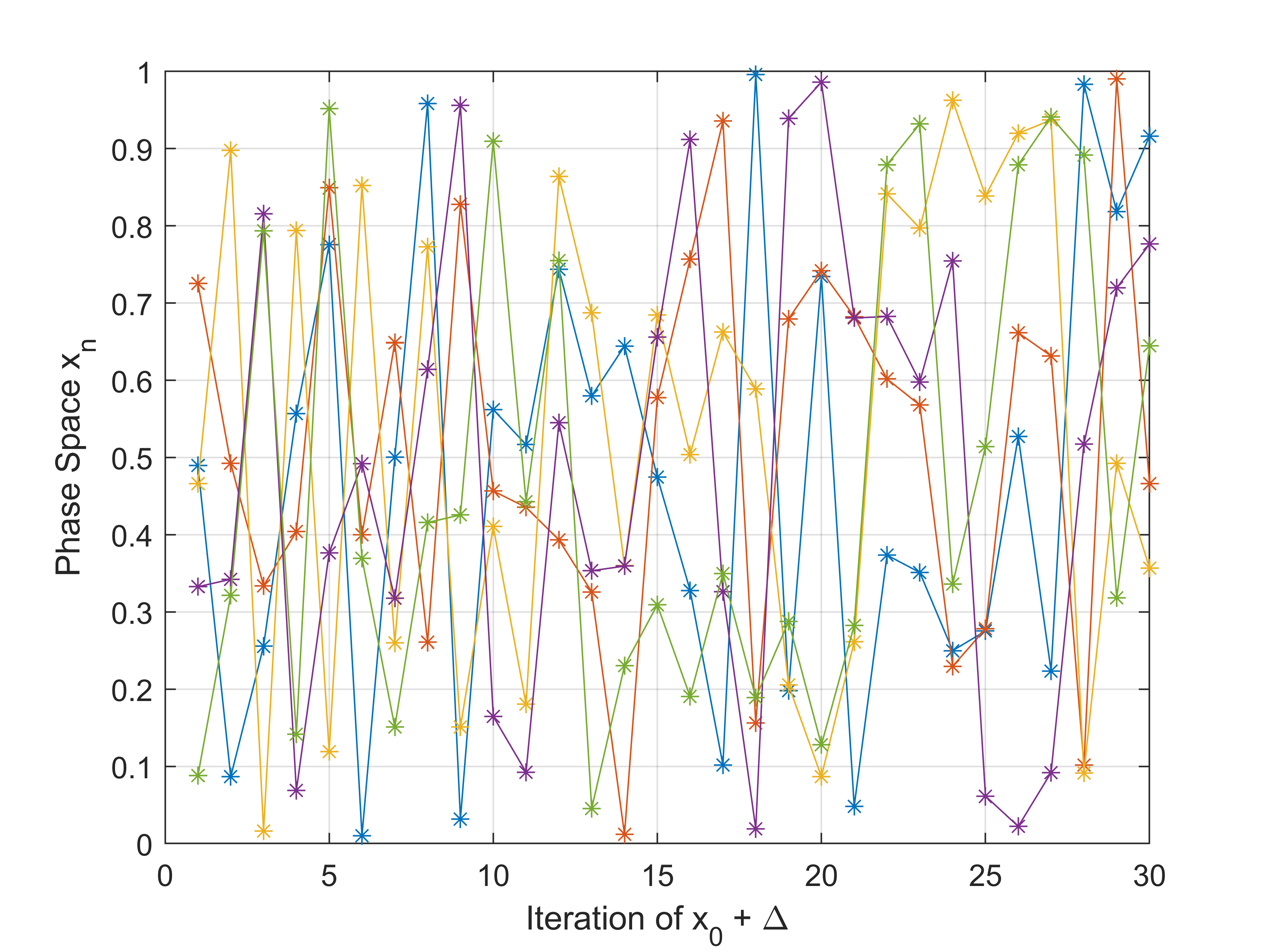}
    \caption{}
    \label{fig:figure_b}
  \end{subfigure}
  \caption{Key Sensitivity under different conditions a) Case 1 and b) Case 2}
  \label{fig:sensitivity_operation}
\end{figure}

Where $\Delta = 2^{-48}$ , in both cases, five sequences of 3000 lengths are generated; only the first 30 values of the sequences are drawn in Fig.~\ref{fig:sensitivity_operation}. It is clear from the plot that trajectories are entirely different even at a tiny change to input. Moreover, a correlation analysis between the generated trajectories is calculated using Eq.~\ref{eq:correaltion_coeff}. It can be noticed from~\ref{tab:correlation} that the correlation between the generated trajectories is very small it means the original signal has no relationship to the other generated signal with tiny differences. 

% \begin{figure}[!ht]
%   \centering
%   \begin{subfigure}{0.9\columnwidth}
%     \includegraphics[width=\columnwidth]{figures/sensitivity_wrt_mu.png}
%     \caption{}
%     \label{fig:image1}
%   \end{subfigure}
%   \hfill
%   \begin{subfigure}{0.9\columnwidth}
%     \includegraphics[width=\columnwidth]{figures/sensitivity_wrt_x.png}
%     \caption{}
%     \label{fig:image2}
%   \end{subfigure}
%   \caption{Key Sensitivity under different conditions: (a) Case 1 and (b) Case 2}
%   \label{fig:ergo_tent_map}
% \end{figure}

\subsection{Differential Analysis}
A unified average changing intensity (UACI) and net pixel change rate (NPCR) are used for differential analysis. The UACI measures the difference of magnitude among trajectories, while NPCR is used to find the number of different elements between $T_1$ and $T_2$, generated with a slightly different initial parameter. The ideal value of UACI and NCPR is 100\%. Following in Eq.~\eqref{eq:uaci} defines UACI and NPCR is determined by Eqs.~\ref{eq:ncpr} and~\ref{eq:s_i} respectively.

\begin{equation}
    UACI = \frac{100}{N} \times \sum_{i=1}^N \left| T_1\left( i \right) - T_2\left( i \right) \right|
    \label{eq:uaci}
\end{equation}

\begin{equation}
    NCPR = \frac{100}{N} \times \sum_{i=1}^N S \left( i\right)
    \label{eq:ncpr}
\end{equation}

\begin{equation}
S\left( i \right) = 
    \begin{cases}
        0, & \text{if } T_1\left( i \right) = T_2\left( i \right)\\
         1,  & \text{if } T_1\left( i \right) \neq T_2\left( i \right)
    \end{cases}
\label{eq:s_i}
\end{equation}

\begin{figure}
  \centering
  \begin{subfigure}{0.45\textwidth}
    \includegraphics[width=\linewidth]{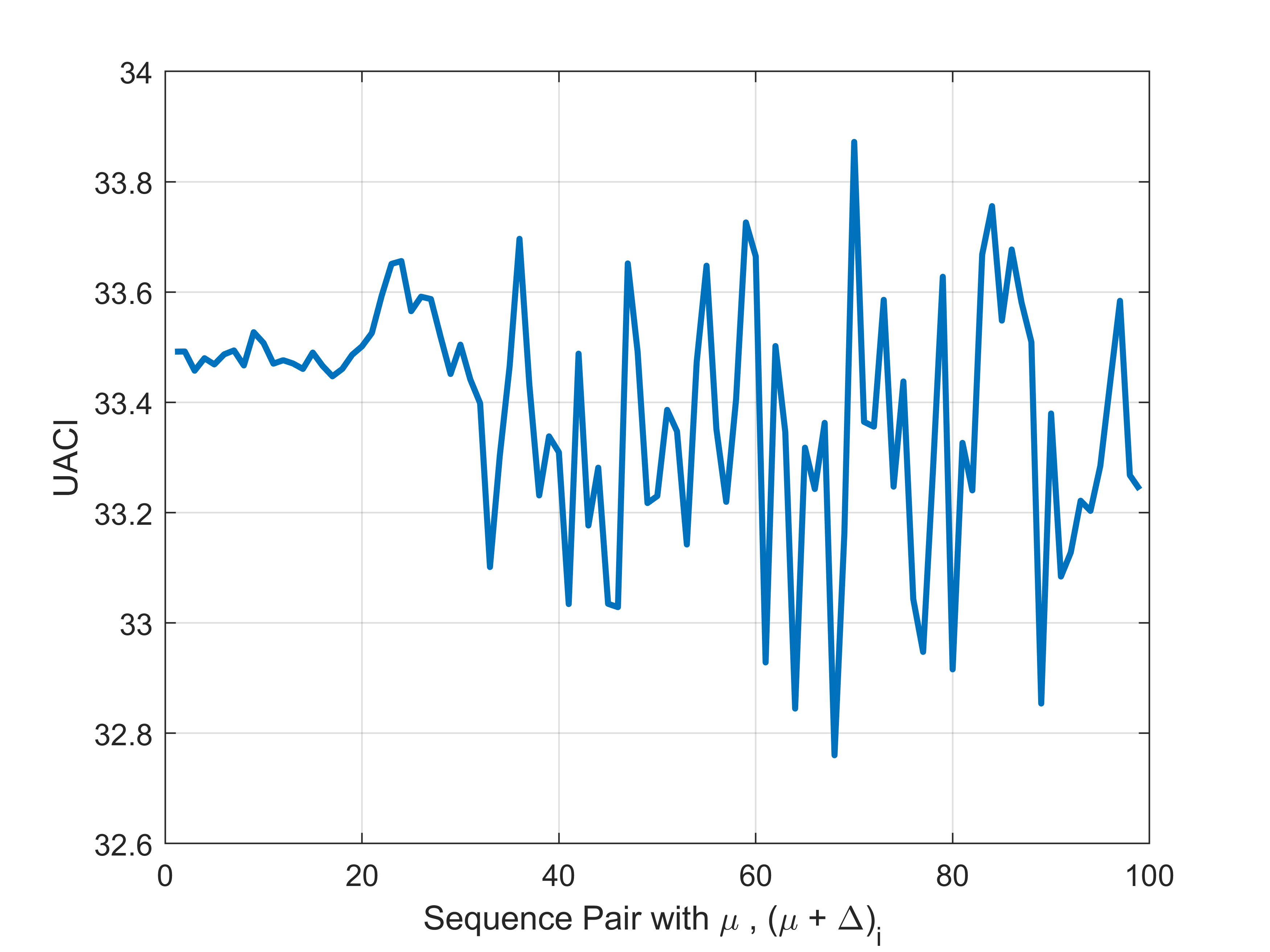}
    \caption{}
    \label{fig:figure_a}
  \end{subfigure}
  \hspace{0.05\textwidth}
  \begin{subfigure}{0.45\textwidth}
    \includegraphics[width=\linewidth]{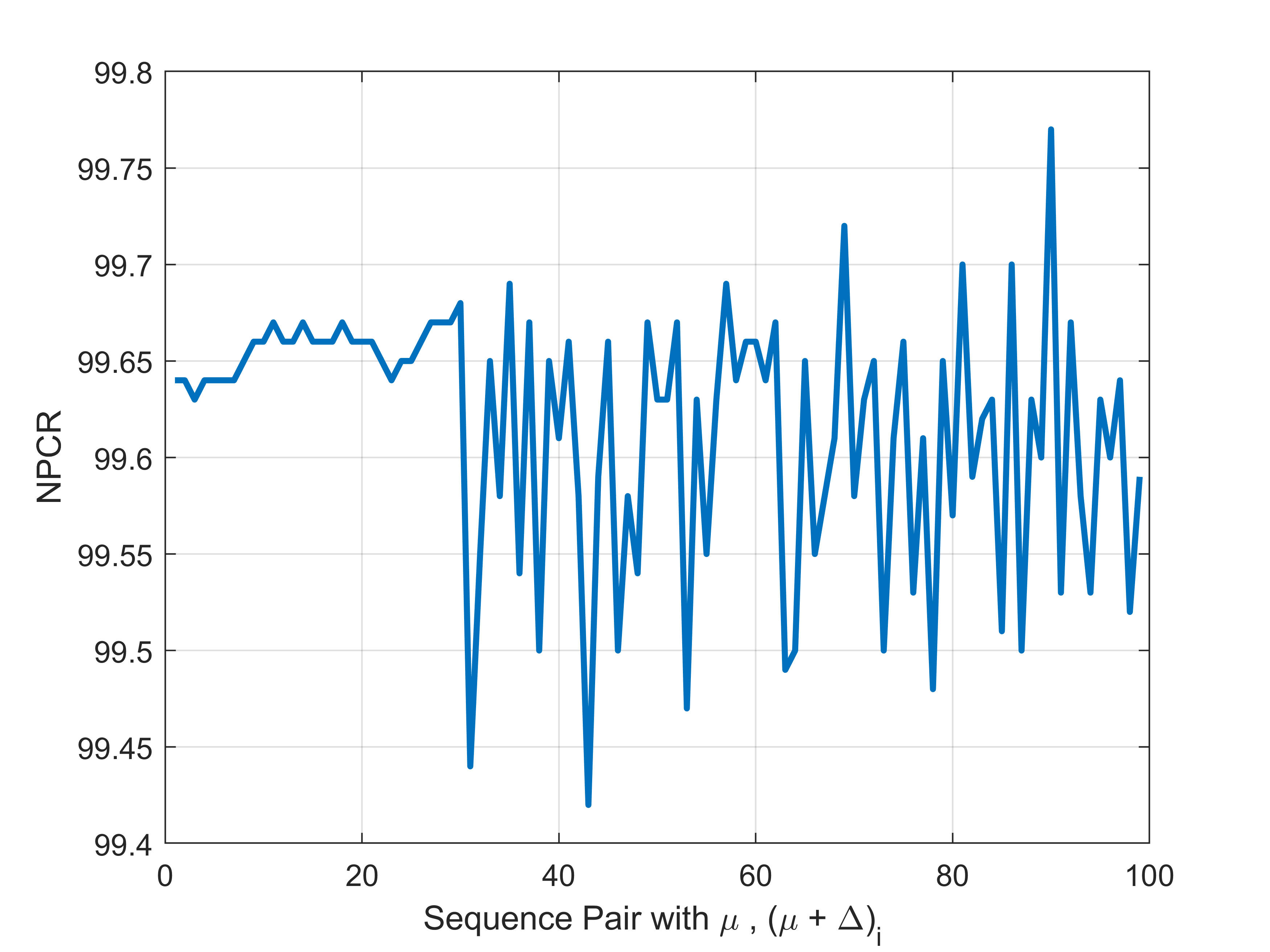}
    \caption{}
    \label{fig:figure_b}
  \end{subfigure}
  \caption{Differential Analysis a) UACI b) NCPR}
  \label{fig:uaci}
\end{figure}

In UACI and NPCR, the same method is applied to sensitivity analysis and generated 100 sequences of size $10^4$, a slight change of $\Delta = 2^{-52}$ is added to $\mu = 93.23$. In 
 Fig.~\ref{fig:uaci} an analysis of NPCR and UACI is presented. The average UACI value of trajectories $T_1$ and $T_2$ is 33.3848\%, and the NPCR mean value is 99.6165\%. Hence, it is concluded that all the trajectories are highly sensitive to the initial conditions; even a small perturbation is applied and produces independent sequences from each other.

\section{Conclusion}
\label{sec:conclusion}
This paper proposed a simple method to design CS-PRBG using RCTM. The RCTM achieves a very larger parameter space. A new set of equations are proposed incorporating modulo and scaling operations.  The generated trajectories are robust, highly nonlinear, and chaotic that can be observed with a positive Lyapunov exponent. The proposed CS-PRBG are analyzed using various statistical methods such as NIST 800-22 test suit, ENT, TestU01, correlation analysis, histogram analysis, key sensitivity analysis, information entropy analysis. The key sensitivity and key-space analysis show that the larger parameter space of the RCTM provides sufficiently larger key space compared to existing PRN generators. The results show that the proposed scheme is suitable for various cryptographic applications and effectively withstand all known attacks.

\section*{Acknowledgment}

Open Access funding provided by the Qatar National Library.
\section*{Disclosure statement}

The authors declare that they have no known competing financial interests or personal relationships that could have appeared to influence the work reported in this paper.
\section*{Data Availability Statement }
The data that support the findings of this study are available from the corresponding author, [author initials], upon reasonable request.
\section*{Authors contributions}
Muhammad Irfan performed conceptualization, formal analysis, funding acquisition, investigation, methodology, software, visualization, validation, writing – original draft, and writing – review \& editing.
Muhammad Asif Khan conducted supervision, validation, writing – original draft, writing – review \& editing, conceptualization, investigation, and methodology.

\bibliographystyle{IEEEtran}
\bibliography{main}

\end{document}